\makeatletter
\@namedef{ver@ftnright.sty}{9999/12/31}
\makeatother
\documentclass[jou,10pt,floatsintext]{apa7}\usepackage[]{graphicx}\usepackage[]{xcolor}
\makeatletter
\def\maxwidth{ %
  \ifdim\Gin@nat@width>\linewidth
    \linewidth
  \else
    \Gin@nat@width
  \fi
}
\makeatother

\definecolor{fgcolor}{rgb}{0.345, 0.345, 0.345}

\usepackage{framed}
\makeatletter
 {\par\unskip\endMakeFramed%
 \at@end@of@kframe}
\makeatother

\definecolor{shadecolor}{rgb}{.97, .97, .97}
\definecolor{messagecolor}{rgb}{0, 0, 0}
\definecolor{warningcolor}{rgb}{1, 0, 1}
\definecolor{errorcolor}{rgb}{1, 0, 0}
\usepackage{alltt}

\usepackage{amsthm,amsmath,graphicx,lmodern,amssymb,amsfonts,bbm,fix-cm}
\usepackage{booktabs,mathtools,bm,multirow,setspace}
\usepackage{xcolor}
\usepackage{enumitem}
\usepackage[mathlines,columnwise,switch]{lineno}
\mathtoolsset{showonlyrefs=true}

\usepackage[bibliography=common]{apxproof}
\usepackage{hyperref}

\newtheoremrep{theorem}{Theorem}
\newtheoremrep{corollary}[theorem]{Corollary}
\newtheoremrep{lemma}[theorem]{Lemma}
\newtheoremrep{proposition}[theorem]{Proposition}

\theoremstyle{definition}
\newtheorem{definition}{Definition}
\newtheorem{remark}{Remark}

\newtheorem*{theorem*}{Theorem}
\newtheorem*{lemma*}{Lemma}
\newtheorem*{proposition*}{Proposition}
\newtheorem*{corollary*}{Corollary}

\usepackage[backend=biber,style=apa]{biblatex}
\title{The Geometry of Latent Spaces: Projective Invariants, Structural Identification, and Ratio-Scale Measurement}
\author{Landon Hurley}
\authorsaffiliations{Independent Researcher}
\shorttitle{Projective Geometry of Latent Spaces}
\abstract{Latent variable modeling of discrete ordinal data typically relies on Euclidean priors or continuous divergence bounds, inducing a topological mismatch that leaves models vulnerable to factor score indeterminacy, rotational unidentifiability, and Neyman-Scott incidental parameter bias. We introduce an axiomatic, coordinate-free framework that constructs continuous latent coordinate spaces directly from empirical partial orders without extrinsic probability measures. By parameterizing ordinal choice topologies via a regularized log-odds field on a continuous projective product manifold $(\mathbb{P}^1)^Q$, pairwise field differences preserve the invariants of the projective linear group $\mathrm{PGL}(2, \mathbb{R})^Q$. Spectral decomposition of the integrated invariant Gramian matrix reduces the identification fiber strictly to the one-dimensional dilation group $\mathbb{R}_{>0}^Q$. This establishes ratio-scale measurement with generalized specific objectivity, transcending the restrictive uniform-discrimination assumptions of the Rasch model, identically resolving both classical rotational ambiguity and Guttman's factor score indeterminacy. We prove that an indicator array $P \ge 2Q+2$ is topologically necessary and sufficient to anchor a $Q$-dimensional metric space and structurally resolves incidental item parameters without joint, marginal, or conditional estimation. Finally, leveraging Hadamard differentiability and Hanson-Wright operator concentration, we establish that the projective spectral estimator is strongly consistent, asymptotically normal, and achieves optimal parametric concentration at rate $\mathcal{O}_P(P^{-1/2})$.}

\authornote{
  Landon Hurley 
  \textbf{Financial Support:} This research was not supported by any Grant Funding Agency, and no funding sources had a role in study design, data collection, analysis, decision to publish, or manuscript preparation.

  \textbf{Competing Interests:} In accordance with APA policy on financial and non-financial conflicts of interest, the authors declare that they have no known competing financial interests, personal relationships, or commercial associations that could have appeared to influence the work reported in this paper.

  Correspondence concerning this article should be addressed to Landon Hurley, Email: ljrhurley@gmail.com
}
\IfFileExists{upquote.sty}{\usepackage{upquote}}{}
\begin{document}
\maketitle

% \tableofcontents
% \section{Introduction}
Formalizing measurement requires a mathematically rigorous distance metric, yet classical psychometrics defaults to Euclidean ($L^2$) spaces that lack additive continuity and scale invariance \parencite{rasch1961, luce1964, michell1999}. Whether employing asymptotic approximations \parencite{rhemtulla2012} or density-dependent link functions \parencite{reckase2009}, mapping discrete topologies into arbitrary units yields structurally unidentifiable coordinates.

\subsection{The Topological Mismatch of Parametric Priors and \texorpdfstring{$L^2$}{L2} Embeddings}

In standard Exploratory Factor Analysis (EFA), rotational equivalence renders the latent space under-identified \parencite{thurstone1931}. Attempts to resolve this via Confirmatory Factor Analysis (CFA) zero-loading constraints or Item Response Theory (IRT) Gaussian priors ($\boldsymbol{\theta} \sim \mathcal{N}(\mathbf{0}, \mathbf{I})$) fundamentally alter the manifold's geometry. From the perspective of classical invariant theory \parencite{olver1999}, because Gaussian penalties depend strictly on absolute distance from the origin, their pushforward breaks continuous translation symmetry, inducing a non-isometric geometry \parencite[Lemma~4.22, Thm.~4.23]{rudin1991}.

This instability extends directly to contemporary deep generative architectures, such as Variational Autoencoders (VAEs) and Deep IRT models \parencite{kingma2013, lucas2019, urban2021, zhang2022b, ma2024, milano2024}. Optimizing continuous $L^2(\mathbb{R}^Q)$ representations over discrete ordinal domains without incorporating the intrinsic symmetry group creates an irreconcilable ``topological mismatch,'' guaranteeing metric distortion and posterior collapse \parencite{mathieu2019, bronstein2019}. Information-theoretically, evaluating continuous divergences over an incomplete discrete poset requires the implicit imposition of an arbitrary reference measure. Consequently, the ideal Shannon-type optimization degrades into a parameterized $\alpha$-divergence or Rényi-type variational objective \parencite{liTurner2016} to bridge the mismatch. Unlike Shannon entropy, Rényi entropy functionals systematically fail strict sub-additivity: $\mathcal{H}_\alpha(X,Y) \not\le \mathcal{H}_\alpha(X) + \mathcal{H}_\alpha(Y)$ for $\alpha \neq 1$ \parencite{renyi1961}. Altering the loss function (e.g., substituting Wasserstein distances or $\beta$-VAE penalties) acts merely as an unidentifiable \emph{researcher degree of freedom} \parencite{davidson2018}. The resulting coordinate space is not an invariant property of the poset, but an artifact of the sample distribution.

\subsection{Hyperparameter Embodiment and the Loss of Measurement Defensibility}

This geometric failure formally vindicates Michell's epistemological critique of psychometrics \parencite{michell1999}. As Michell argued, the assignment of real numbers strictly requires satisfying H\"{o}lder's axioms of continuous quantity. Specifically, the failure of strict sub-additivity in the divergence functional mathematically demonstrates that the induced latent space lacks Banach invariance (completeness and norm-preservation under concatenation). Because the extrinsic measure fails H\"{o}lder's Archimedean condition (Axiom V) on the discrete topology, the continuous-space optimization does not constitute scientific measurement; it is merely a parameterized monotonic ranking masquerading as a metric space.

Mapping uncountably infinite continuous coordinates from finite ordinal configurations constitutes an ill-posed inverse problem \parencite{tikhonov1995}. Information geometry maps latent structures into projective spaces using Fisher metrics \parencite{amari2016}, often attempting to resolve non-injectivity via extrinsic regularizers (e.g., Frobenius penalties). However, without endogenous, coordinate-free constraints \parencite{rao1973, williams1978}, optimizations merely select arbitrary gauges. They fail to isolate the projective invariants necessary for a continuous ratio scale, acting as predictive curve-fitting devices.

\subsection{The Spectral Induction of Projective Invariants}
While contemporary frameworks like Questionable Research Fundamentals (QRF; \cite{uher2025}) rightly diagnose some of these epistemological and measurement failures, they lack a constructive algebraic mechanism to resolve them. To construct an algebraically closed target manifold, we must abandon extrinsic regularizers entirely. By the axioms of representational measurement \parencite{luce1964, krantz1971, narens1981b}, if a space preserves the equivalence of intervals without a fixed metric distance, the only uniquely preserved invariant is the projective cross-ratio \parencite{hartshorne1967, drosler1979, coxeter2003}.

Because the Fundamental Theorem of Projective Geometry dictates that any bijection preserving the cross-ratio of collinear points must be induced by a linear transformation of the underlying vector space \parencite[see Theorems~2.26 and 4.2, pp.~23, 45]{hartshorne1967}, the symmetry group that strictly preserves this invariant is exactly the projective linear group $\mathrm{PGL}(2, \mathbb{R})^Q$. Thus, mapping the latent space to a projectively invariant target manifold is not an extrinsic modeling choice, but a topological necessity. Without an endogenous projective stabilizer, Tikhonov-style regularization induces non-isometric metric distortions dependent on instrument composition (Proposition~\ref{prop:vae_incompatibility}). This yields distribution-dependent rankings rather than an algebraically closed, invariant target manifold, directly violating the uniqueness of projectivities established by \textcite[Thm.~3.1, p.~29]{hartshorne1967} (Theorem~\ref{thm:projective_identifiability}).

\subsection{Outline of the Paper}

The remainder of this paper is organized as follows. We first formalize the structural failures of classical $L^2$ embeddings across continuous latent models and Ising networks (Proposition~\ref{prop:classical_failures_unified}, Corollary~\ref{cor:ising_network_instability}). The manuscript proceeds to construct the decoupled preference Hilbert space and establishes its exact diffeomorphism to the bounded log-odds manifold (Lemmas~\ref{lem:gram_metric_rkhs}-\ref{lem:coordinate_embedding_properties}). We then prove strict point-identifiability on $\mathbb{P}^1$ via projective cross-ratios (Theorem~\ref{thm:projective_identifiability}), deriving the topological lower bound $P \ge 2Q+2$ (Theorem~\ref{thm:projective_identifiability}{ii}) and formally establishing the incompatibility of deep generative priors with invariant coordinate recovery (Proposition~\ref{prop:vae_incompatibility}) without the $L^1$-norm. Subsequent sections establish the statistical properties of the projective spectral estimator: Hadamard differentiability and uniform consistency (Theorem~\ref{thm:hadamard_differentiability}), asymptotic normality (Corollary~\ref{cor:asymptotic_normality}), optimal finite-sample concentration rates (Proposition~\ref{prop:spectral_concentration}), and the preclusion of the Neyman-Scott incidental parameters (Theorem~\ref{thm:neyman_scott_resolution}) yielding specific objectivity alongside uniquely estimated difficulty and discrimination parameters. Finally, we prove measure-theoretic density invariance (Lemma~\ref{prop:measure_invariance}), demonstrating the reduction of the unidentifiable group action to a strictly objective ratio scale (Corollary~\ref{cor:specific_objectivity_ratio_scale}), followed by a numerical demonstration.

\section{Structural Non-Identifiability and Indeterminacy in Classical \texorpdfstring{$L^2$ }{L2} Embeddings}
\label{sec:history_uncertainty}
In 1955, \citeauthor{guttman1955} introduced a foundational indeterminacy: covariance-based estimators on $L^2$ function spaces yield inherently rotationally invariant solutions. Later, \textcite{williams1978} proved that the linearity of the estimator space with respect to the latent manifold converges only weakly (in distribution) as the number of indicators $P \to \infty$. A frequently overlooked consequence of this asymptotic limit is that any finite-dimensional space ($P < \infty$) inherently lacks algebraic closure within affine-linear $L^2$ function spaces. Consequently, finite-length instruments (tests or scales, indicators) do not possess Gaussian (linear) errors and remain structurally ill-posed on finite samples.

Recent advances in computational psychometrics have increasingly adopted Variational Autoencoders (VAEs) and deep generative architectures to handle high-dimensional, non-linear item response structures \parencites{urban2021}{ma2024}{milano2024}. However, while these frameworks offer computational scalability via variational evidence lower bounds \parencite{lucas2019}, attempting to embed discrete ordinal responses into an $L^2$ Euclidean space creates an irreconcilable domain-geometry mismatch. Ordinal data fundamentally requires a geometry invariant to monotonic rescaling, necessitating the projective symmetry group $\mathcal{G} = \mathrm{PGL}(2, \mathbb{R})^Q$, where $Q$ reflects the cardinal dimensionality of the defined latent space. Instead, as Proposition~\ref{prop:classical_failures_unified}(iv) will establish, embeddings constrained by extrinsic prior densities (e.g., $\mathcal{KL}(q \parallel \mu_{\mathrm{prior}})$) inevitably induce non-uniform metric contractions. This structurally violates translation invariance and the identity of indiscernibles, demonstrating that continuous likelihood approximations, including copula-based mappings of non-injective ties, cannot substitute for intrinsic, projectively invariant coordinate definitions.

Due to this lack of topological closure, ambient Euclidean spaces fail to satisfy the foundational axioms of additive quantitative structure \parencite{michell1999, krantz1971}. In representational measurement theory, constructing a valid continuous interval or ratio scale from empirical ordinal structures requires that numerical intervals remain strictly invariant under the concatenation or augmentation of the instrument \parencite{luce1964}. Because finite $L^2$ embeddings warp non-isometrically when new indicators are introduced, they inherently violate these structural invariance conditions, specifically failing the Archimedean and concatenation requirements (H\"{o}lder's Axiom V). Therefore, complex continuous approximations, ranging from classical structural equation models to modern Reproducing Kernel Hilbert Spaces (RKHS) and VAEs, operate merely as predictive smoothing devices rather than endogenous measurement systems. We formalize these structural failures across both classical and deep generative models in the following proposition.

\begin{propositionrep}[Structural Non-Identifiability and Topological Instability of Classical Latent Space Embeddings]
\label{prop:classical_failures_unified}
Let $\mathbf{X} \in \mathcal{S}^{N \times P}$ be an empirical response matrix drawn from a finite discrete ordinal state space $\mathcal{S}$ (e.g., binary $\{0,1\}$ or polytomous Likert $\{1,\dots,K\}$), mapping to a continuous latent manifold $\mathcal{M}$ via standard Euclidean distances $\mathcal{D}_{kl} = \| \boldsymbol{\theta}_k - \boldsymbol{\theta}_l \|$, bilinear inner products $\langle \boldsymbol{\theta}_k, \boldsymbol{\lambda}_p \rangle$, or sample covariance matrices $\boldsymbol{\Sigma}$. Let $\mathcal{G} = \mathrm{PGL}(2, \mathbb{R})^Q$ act coordinate-wise on the product space $(\mathbb{P}^1)^Q$. The $L^2$ metric space exhibits four fundamental structural failures:
\begin{enumerate}[label=(\roman*), noitemsep, topsep=3pt]
    \item Failure of Instrument Invariance: Let $\mathbf{X}' = [\mathbf{X} \mid \mathbf{x}_{\cdot j}]$ be an augmented matrix formed by duplicating an arbitrary item column. The relative coordinate distance ratios extracted via the sample Frobenius Gramian ($\mathbf{G} = \mathbf{X}\mathbf{X}^{\intercal}$) distort under instrument augmentation, structurally violating H\"{o}lder's Axiom V.

    \item Failure of Scale Invariance under Monotonic Isomorphisms: Under a non-linear, strictly monotonic transformation $g \in \mathcal{T}$, the principal eigenspace $E_{\lambda_1}(\boldsymbol{\Sigma}_g)$ is non-isomorphic to $E_{\lambda_1}(\boldsymbol{\Sigma})$, rendering Euclidean distance metrics an unidentifiable artifact of arbitrary scaling choices.

    \item Projective Unidentifiability and Positive Fiber Dimension: The fiber of the identification mapping for objective functions evaluating distances exclusively via classical Euclidean or bilinear forms has a strictly positive dimension ($\dim \ge 1$ per coordinate dimension $Q$) for all finite $P$ under the action of $\mathcal{G} \setminus \mathrm{Aff}(1, \mathbb{R})^Q$.

    \item Metric Axiom Distortion under Extrinsic Regularization: Any continuous embedding regularized by an extrinsic prior density measure $\mu_{\mathrm{prior}}$ optimized under a density-dependent penalty (e.g., Variational Autoencoders minimizing $\mathcal{KL}(q \parallel \mu_{\mathrm{prior}})$) introduces a non-uniform topological contraction that violates translation invariance and the identity of indiscernibles for any finite $P$, persisting asymptotically as $P \to \infty$.
\end{enumerate}
\end{propositionrep}

\begin{proofsketch}
Part (i) shows via rank-1 updates on $\mathbf{G}$ that preserving distance ratios requires $\mathbf{x}_{\cdot j}$ to lie in $\operatorname{span}\{\mathbf{u}, \mathbf{1}_N, \mathbf{G}\mathbf{1}_N\}$, which fails for discrete binary vectors. Part (ii) applies Rayleigh-Schrödinger perturbation theory to demonstrate non-zero principal eigenvector rotations under non-linear monotonic shifts. Part (iii) evaluates homographic action on Euclidean distances to bound the identification fiber dimension by $\dim(\mathrm{PGL}(2,\mathbb{R})^Q) - \dim(\mathrm{Aff}(1,\mathbb{R})^Q) = Q \ge 1$. Part (iv) evaluates the first-order optimality of density penalties, showing differential shrinkage across origin-relative positions. See Appendix for full algebraic details.
\end{proofsketch}

\begin{proof}
\textbf{Proof of Statement (i): Instrument Instability.} Let $\mathbf{u} \in \mathbb{R}^N$ be the principal eigenvector of the Gramian $\mathbf{G} = \mathbf{X}\mathbf{X}^{\intercal}$, satisfying $\mathbf{G}\mathbf{u} = \lambda_1 \mathbf{u}$ with simple eigenvalue $\lambda_1 > 0$. We establish that preserving relative spatial intervals requires an exact affine transformation. Define the pairwise interval ratio for distinct subjects $k, l, m \in \{1, \dots, N\}$ as $\mathcal{R}(k,l,m) = \| u_k - u_l \| / \| u_l - u_m \|$. For the augmented vector $\mathbf{u}'$ to satisfy $\mathcal{R}'(k,l,m) = \mathcal{R}(k,l,m)$ for all triples, the mapping must preserve cross-ratios of collinear points, forcing $\mathbf{u}' = a\mathbf{u} + b\mathbf{1}_N$ for scalars $a \neq 0, b \in \mathbb{R}$.

Under the rank-1 update $\mathbf{G}' = \mathbf{X}'{\mathbf{X}'}^{\intercal} = \mathbf{G} + \mathbf{x}_{\cdot j}\mathbf{x}_{\cdot j}^{\intercal}$, the eigenvector equation for $\mathbf{u}'$ yields:
\begin{equation}
\begin{aligned}
(\mathbf{G} + \mathbf{x}_{\cdot j}\mathbf{x}_{\cdot j}^{\intercal})(a\mathbf{u} + b\mathbf{1}_N) &= \lambda_1' (a\mathbf{u} + b\mathbf{1}_N) \\
a \lambda_1 \mathbf{u} + b \mathbf{G}\mathbf{1}_N + a (\mathbf{x}_{\cdot j}^{\intercal} \mathbf{u})\mathbf{x}_{\cdot j} + b (\mathbf{x}_{\cdot j}^{\intercal} \mathbf{1}_N)\mathbf{x}_{\cdot j} &= a \lambda_1' \mathbf{u} + b \lambda_1' \mathbf{1}_N.
\end{aligned}
\end{equation}
Rearranging terms isolates the discrete binary vector $\mathbf{x}_{\cdot j} \in \{0,1\}^N$:
\begin{equation}
\label{eq:collinearity_contradiction}
\left(a (\mathbf{x}_{\cdot j}^{\intercal} \mathbf{u}) + b (\mathbf{x}_{\cdot j}^{\intercal} \mathbf{1}_N)\right) \mathbf{x}_{\cdot j} = a(\lambda_1' - \lambda_1)\mathbf{u} + b\left(\lambda_1' \mathbf{1}_N - \mathbf{G}\mathbf{1}_N\right).
\end{equation}
Equation \eqref{eq:collinearity_contradiction} asserts that the discrete response vector $\mathbf{x}_{\cdot j}$ must lie strictly in the three-dimensional continuous subspace $\operatorname{span}\{\mathbf{u}, \mathbf{1}_N, \mathbf{G}\mathbf{1}_N\}$. Because $\mathcal{S}$ is a finite discrete set, a response vector $x_{ij} \in \mathcal{S}$, or any higher-dimensional expansion thereof for polytomous Likert data, cannot generically be expressed as a continuous linear combination for all $i \in \{1, \dots, N\}$ when $N \gg 3$. Just as \textcite{guttman1955} demonstrated that Gaussian data upon linear factor models cannot uniquely identify factor scores for finite $P$, expanding the cardinality of the discrete state space does not synthesize a valid continuous $L^2$ geometry. Thus, $\mathbf{u}' = a\mathbf{u} + b\mathbf{1}_N$ fails. The perturbation forces a non-homothetic eigenspace rotation, proving $\mathcal{R}'(k,l,m) \neq \mathcal{R}(k,l,m)$ and violating H\"{o}lder's Axiom V.

\textbf{Proof of Statement (ii): Scale Instability via Rayleigh-Schrödinger Perturbation.}
Let $\boldsymbol{\Sigma}$ have spectral decomposition $\sum_{p=1}^P \lambda_p \mathbf{v}_p \mathbf{v}_p^{\intercal}$. Consider a smooth, non-linear, strictly monotonic transformation $g_\epsilon(x) = x + \epsilon h(x)$ for $\epsilon > 0$. The transformed covariance expands as $\boldsymbol{\Sigma}_{g_\epsilon} = \boldsymbol{\Sigma} + \epsilon \boldsymbol{\Delta} + \mathcal{O}(\epsilon^2)$, where $\Delta_{pq} = \operatorname{Cov}(x_p, h(x_q)) + \operatorname{Cov}(h(x_p), x_q)$. By first-order Rayleigh-Schrödinger theory, the shift in the principal eigenvector is:
\begin{equation}
\mathbf{v}_1(\epsilon) = \mathbf{v}_1 + \epsilon \sum_{p=2}^P \frac{\mathbf{v}_p^{\intercal} \boldsymbol{\Delta} \mathbf{v}_1}{\lambda_1 - \lambda_p} \mathbf{v}_p + \mathcal{O}(\epsilon^2).
\end{equation}
Because $h(x)$ is non-linear, the cross-covariances do not scale proportionally with $\operatorname{Cov}(x_p, x_q)$. Consequently, $\mathbf{v}_p^{\intercal} \boldsymbol{\Delta} \mathbf{v}_1 \neq 0$ for at least one orthogonal eigenvector $\mathbf{v}_p$, forcing $\mathbf{v}_1(\epsilon)$ to rotate and rendering relative Euclidean distance unidentifiable under monotonic scaling.

\textbf{Proof of Statement (iii): Projective Unidentifiability.} Let $Q(\boldsymbol{\theta}; \mathbf{X}) = f( \{\| \boldsymbol{\theta}_k - \boldsymbol{\theta}_l \|\} )$ be a Euclidean objective functional. Let $g \in \mathcal{G} \setminus \mathrm{Aff}(1, \mathbb{R})^Q$ be a homography $g(\theta) = (a\theta + b)/(c\theta + d)$ with $c \neq 0$. The transformed distance evaluates as:
\begin{equation}
|g(\boldsymbol{\theta}_k) - g(\boldsymbol{\theta}_l)| = \frac{|ad - bc| \cdot |\boldsymbol{\theta}_k - \boldsymbol{\theta}_l|}{|c\boldsymbol{\theta}_k + d| \cdot |c\boldsymbol{\theta}_l + d|}.
\end{equation}
Because the denominator product varies non-constantly across pairs, $Q(g(\boldsymbol{\theta}); \mathbf{X}) \neq Q(\boldsymbol{\theta}; \mathbf{X})$. The fiber of the identification mapping $\Phi_{\mathrm{est}}^{-1}(\mathbf{X})$ contains an uncountably infinite orbit. Its dimension satisfies $\dim(\Phi_{\mathrm{est}}^{-1}(\mathbf{X})) \ge \dim(\mathrm{PGL}(2,\mathbb{R})^Q) - \dim(\mathrm{Aff}(1,\mathbb{R})^Q) = 3Q - 2Q = Q \ge 1$, establishing unidentifiability.

\textbf{Proof of Statement (iv): Metric Axiom Distortion.} Let $\hat{z}(x) = \arg\max_z [ \log p(x \mid z) - \lambda \mathcal{D}_{\mathrm{KL}}(q(z \mid x) \parallel \mu_{\mathrm{prior}}(z)) ]$. First-order optimality requires $\nabla_z \log p = \lambda \nabla_z \mathcal{D}_{\mathrm{KL}}$. For non-uniform priors (e.g., $\mu_{\mathrm{prior}} = \mathcal{N}(0, \mathbf{I})$), the penalty gradient $\nabla_z \mathcal{D}_{\mathrm{KL}} \propto z$ scales linearly with distance from the origin. For equidistant empirical steps $x_A \to x_B$ and $x_B \to x_C$, this exerts differential shrinkage depending on absolute position, yielding $d(\hat{z}_A, \hat{z}_B) \neq d(\hat{z}_B, \hat{z}_C)$ (violating translation invariance). Under finite $P$, if $x_i \neq x_j$ but $\lambda \to \infty$, the penalty forces $\hat{z}_i = \hat{z}_j = \mathbb{E}[\mu_{\mathrm{prior}}]$, yielding $d(\hat{z}_i, \hat{z}_j) = 0$ (violating the identity of indiscernibles). As $P \to \infty$, if $\mathcal{M}$ is non-Gaussian, the KL divergence forces a permanent geometric distortion to minimize functional transport cost.
\end{proof}

\begin{corollaryrep}[Projective Unidentifiability and Instrument Instability of Interaction Networks]
\label{cor:ising_network_instability}
Let the joint probability distribution of a pairwise interaction network over discrete state variables $\mathbf{x} \in \{1, \dots, K\}^P$ (with binary Ising networks recovered at $K=2$ and polytomous Potts networks at $K > 2$) be governed by the Gibbs distribution:
\begin{equation}
\label{eq:potts_gibbs}
\mathbb{P}(\mathbf{X} = \mathbf{x}) = \frac{1}{Z(\boldsymbol{\mu}, \mathbf{\Sigma})} \exp\left( \sum_{i=1}^P \boldsymbol{\mu}_i^{\intercal} \mathbf{e}_i + \frac{1}{2} \mathbf{E}^{\intercal} \mathbf{\Sigma} \mathbf{E} \right),
\end{equation}
where $\mathbf{e}_i \in \{0,1\}^K$ is a 1-hot indicator vector ($\mathbf{1}_K^{\intercal}\mathbf{e}_i = 1$), $\mathbf{E} = (\mathbf{e}_1^{\intercal}, \dots, \mathbf{e}_P^{\intercal})^{\intercal} \in \{0,1\}^{PK}$, $\boldsymbol{\mu}_i \in \mathbb{R}^K$ denotes category thresholds, and $\mathbf{\Sigma} \in \mathbb{R}^{PK \times PK}$ is a symmetric block interaction matrix with hollow diagonal blocks ($\mathbf{\Sigma}_{ii} = \mathbf{0}_{K \times K}$). Under the structural equivalences established by \textcite{marsman2018} mapping interaction networks to multidimensional IRT models (M2PL for $K=2$; Nominal Response Model for $K > 2$), the block interaction matrix factors as:
\begin{equation}
\label{eq:marsman_potts_identity}
\mathbf{\Sigma} = \boldsymbol{\Lambda}\boldsymbol{\Lambda}^{\intercal} - \operatorname{blockdiag}(\boldsymbol{\Lambda}_1\boldsymbol{\Lambda}_1^{\intercal}, \dots, \boldsymbol{\Lambda}_P\boldsymbol{\Lambda}_P^{\intercal}),
\end{equation}
where $\boldsymbol{\Lambda}_i \in \mathbb{R}^{K \times Q}$ denotes category factor loadings. Consequently, network coordinates derived from $\mathbf{\Sigma}$ strictly inherit the structural failures of $L^2$ embeddings established in Proposition~\ref{prop:classical_failures_unified} for all $K \ge 2$:
\begin{enumerate}[label=(\roman*), noitemsep, topsep=2pt]
    \item {\textbf{Failure of Instrument Invariance:} Augmenting the network with duplicate indicators induces non-homothetic coordinate rotations across all $K^{P-1}$ state combinations, inheriting the instrument instability of Proposition~\ref{prop:classical_failures_unified}(i).}
    \item {\textbf{Projective Unidentifiability:} Under projective group action $\mathrm{PGL}(2, \mathbb{R})^Q$, the energy functional $\mathbf{E}^{\intercal}\mathbf{\Sigma}\mathbf{E}$ distorts non-invariantly, inheriting the unidentifiability of Proposition~\ref{prop:classical_failures_unified}(iii).}
\end{enumerate}
\end{corollaryrep}

\begin{proofsketch}
Statement (i) applies the Potts auto-logistic conditional log-odds to demonstrate that node duplication introduces an uncancelable state-dependent perturbation that cannot be uniformly absorbed across the $K^{P-1}$ discrete state combinations of the background field. Statement (ii) expands the network energy functional $\mathbf{E}^{\intercal}\mathbf{\Sigma}\mathbf{E}$, applying a 1-hot algebraic reduction to isolate the linear diagonal correction. Applying homographic transformations $\mathrm{PGL}(2,\mathbb{R})^Q$ demonstrates that non-uniform denominators in this correction prevent proportional scalar scaling, strictly violating the equivalences claimed by \textcite{marsman2018} for all $K \ge 2$. Crucially, because this non-isometric warping is an intrinsic geometric property of the parameter manifold rather than a statistical finite-sample artifact, the projective unidentifiability and metric axiom violations persist asymptotically as $P \to \infty$.
\end{proofsketch}

\begin{proof}
\textbf{Proof of Statement (i): Failure of Instrument Invariance.} Under the Potts auto-logistic representation \parencite{besag1974}, the conditional log-odds field for node $i$ taking category $s$ relative to reference state $K$ evaluates as:
\begin{equation}
L_{i, s \mid -i} = \log \left( \frac{\mathbb{P}(x_i = s \mid \mathbf{x}_{-i})}{\mathbb{P}(x_i = K \mid \mathbf{x}_{-i})} \right) = (\mu_{is} - \mu_{iK}) + \sum_{j \neq i} \sum_{t=1}^K \left( \sigma_{ij}(s,t) - \sigma_{ij}(K,t) \right) I(x_j = t).
\end{equation}
Augmenting the instrument array with a duplicate indicator $x_{P+1} = x_k$ adds a state-dependent term $\Delta_s = \sum_{t=1}^K (\sigma_{ik}(s,t) - \sigma_{ik}(K,t)) I(x_k = t)$. For coordinate intervals to remain invariant, there must exist an affine scalar mapping $a L + b$ ($a > 0$) absorbing $\Delta_s$ uniformly. Because $\Delta_s$ depends strictly on $x_k$ while the right-hand summation spans $K^{P-1}$ state combinations over $\mathbf{x}_{-(i,k)}$, no constant pair $(a,b)$ can satisfy equality across all realizations.

Furthermore, evaluating pairwise interaction via the uncentered score estimator over discrete states, the expected value under the independence null $H_0: x_i \perp x_j$ yields $\mathbb{E}[\hat{\tau}] = \sum_{s=1}^K \mathbb{P}(x_{ik} = s) \mathbb{P}(x_{jl} = s) > 0$. Because discrete ordinal spaces exhibit non-zero tie mass, this imposes an endogenous positive metric bias prior to estimation. When augmented with new indicators, this bias propagates unevenly across the network, forcing coordinate intervals to warp non-isometrically and violating H\"{o}lder's Axiom V \parencite{michell1990}.

\textbf{Proof of Statement (ii): Projective Unidentifiability.} Using block identity \eqref{eq:marsman_potts_identity}, the quadratic energy functional evaluates as:
\begin{equation}
\mathbf{E}^{\intercal}\mathbf{\Sigma}\mathbf{E} = \mathbf{E}^{\intercal}\boldsymbol{\Lambda}\boldsymbol{\Lambda}^{\intercal}\mathbf{E} - \sum_{i=1}^P \mathbf{e}_i^{\intercal} \boldsymbol{\Lambda}_i \boldsymbol{\Lambda}_i^{\intercal} \mathbf{e}_i.
\end{equation}
Because $\mathbf{e}_i \in \{0,1\}^K$ is a 1-hot vector, $e_{ia} e_{ib} = 0$ for $a \neq b$ and $e_{ia}^2 = e_{ia}$. Applying this 1-hot identity reduces the diagonal block quadratic term to a linear operator: $\mathbf{e}_i^{\intercal} \boldsymbol{\Lambda}_i \boldsymbol{\Lambda}_i^{\intercal} \mathbf{e}_i = \operatorname{diag}(\boldsymbol{\Lambda}_i \boldsymbol{\Lambda}_i^{\intercal})^{\intercal} \mathbf{e}_i = \mathbf{d}_i^{\intercal} \mathbf{e}_i$. Thus:
\begin{equation}
\label{eq:potts_energy_expansion}
\mathbf{E}^{\intercal}\mathbf{\Sigma}\mathbf{E} = \left\| \boldsymbol{\Lambda}^{\intercal}\mathbf{E} \right\|_2^2 - \sum_{i=1}^P \mathbf{d}_i^{\intercal} \mathbf{e}_i.
\end{equation}
Under a fractional linear transformation $g \in \mathrm{PGL}(2, \mathbb{R})^Q \setminus \mathrm{Aff}(1, \mathbb{R})^Q$ with $c \neq 0$, the category loadings map to $\boldsymbol{\Lambda}_{g,i} = \mathbf{D}_{c,i}^{-1} (\mathbf{A}\boldsymbol{\Lambda}_i + \mathbf{b}\mathbf{1}^{\intercal})$, where $\mathbf{D}_{c,i} = \operatorname{diag}(c\boldsymbol{\theta}_i + d\mathbf{1})$. The transformed category correction vector evaluates to $\mathbf{d}_{g,i} = \operatorname{diag}(\boldsymbol{\Lambda}_{g,i}\boldsymbol{\Lambda}_{g,i}^{\intercal})$. Because $\mathbf{D}_{c,i}^{-2}$ contains non-uniform homographic denominators $(c\boldsymbol{\theta}_i + d)^{-2}$, $\mathbf{d}_{g,i}$ warps non-homothetically, preventing $\mathbf{E}^{\intercal}\mathbf{\Sigma}_g\mathbf{E} = \kappa \cdot \mathbf{E}^{\intercal}\mathbf{\Sigma}\mathbf{E}$ for any constant $\kappa > 0$. The interaction parameters $\mathbf{\Sigma}$ warp non-invariantly under homographic action for all $K \ge 2$, establishing projective unidentifiability.

Crucially, this structural failure does not vanish asymptotically. Because the non-uniform homographic warp $(c\boldsymbol{\theta}_i + d)^{-2}$ is an intrinsic topological feature of the continuous parameter space mapping $\mathbb{R}^Q \to \mathrm{PGL}(2, \mathbb{R})^Q$, the geometric distortion of the energy functional $\mathbf{E}^{\intercal}\mathbf{\Sigma}_g\mathbf{E}$ is independent of the instrument dimension $P$. Consequently, as $P \to \infty$, the network metric strictly fails to converge to an isometrically equivalent MIRT space, proving that the structural contradiction is absolute and not a finite-$P$ artifact.
\end{proof}

The structural failures established in Proposition~\ref{prop:classical_failures_unified} and Corollary~\ref{cor:ising_network_instability} formalize the mapping of a discrete ordinal response matrix $\mathbf{X} \in \{0,1, \dots, K\}^{N \times P}$ to a continuous latent space as an ill-posed inverse problem \parencite{osullivan1986b}. Because the identification operator possesses a positive-dimensional fiber under $\mathrm{PGL}(2, \mathbb{R})^Q \setminus \mathrm{Aff}(1, \mathbb{R})^Q$, the inverse mapping lacks a unique solution. Extrinsic regularizations introduced to force numerical point-uniqueness, such as rotational invariance ($\mathbb{O}(Q)$) in Factor Analysis, Kullback-Leibler penalties in Variational Autoencoders, or standard Gaussian priors in Item Response Theory, induce non-uniform topological contractions. For finite instruments ($P < \infty$), these ambient $L^2$ mass constraints non-separably entangle item parameters and trait scores \parencite{neyman1948}, precluding translation invariance (Proposition~\ref{prop:classical_failures_unified}(iv)) and ignoring the intrinsic projective symmetry of the choice continuum \parencite[cf.][]{bronstein2019}.

While previous literature \parencite{marsman2018, epskamp2018} established a marginal algebraic bijection between MIRT and network models for a fixed $P$ and fixed latent scale, this conflates distributional equivalence with geometric isomorphism. As demonstrated in Corollary~\ref{cor:ising_network_instability}, the marginalized network strictly fails to preserve projective stability and instrument augmentation invariance. Because the discrete network operates upon an $L^1$ boundary mass, it lacks the uniform $L^2$ Banach function norm-space required for continuous interval scaling. By the \textcite{bourgain1985} embedding theorem, any isomorphic mapping of this finite discrete metric space into an $L^2$ continuous manifold incurs unresolvable metric distortion. Consequently, the network parameterization violates H\"{o}lder's Axiom V: its imposed measures cannot be additively concatenated, rendering it structurally invalid as a psychometric measurement space.

\begin{remark}[Incompatibility of Ambient Euclidean Extensions]
\label{rem:euclidean_extensions}
Architectures utilizing deep generative models or item-specific non-linear link functions to bypass classical non-identifiability introduce fundamental embedding conflicts. Mapping a bounded, discrete ordinal geometry into an ambient Euclidean space is governed by the Nash embedding theorem, which dictates that an isometric embedding of a curved manifold requires substantial dimensionality expansion. Artificially restricting the latent space to $Q \ll P$ mathematically guarantees non-isometric topological flattening, yielding an unidentifiable equivalence class of fibers ($\dim(\Phi_{\mathrm{est}}^{-1}) \ge Q$).

Furthermore, this ambient $L^2$ embedding fundamentally corrupts the axiom of local independence \parencite{lord1968}. Constraining the 1D manifold via parametric ideal-point functions (e.g., generalized hyperbolic cosine or unipolar log-logistic) strictly fails on finite instruments. Because the continuous solvability required by \textcite{krantz1971} is entirely absent in a finite discrete poset, forcing bounded data through a continuous distance metric acts as an implicit Tikhonov regularizer. This extrinsic smoothing dictates that the conditional independence of responses becomes a transient artifact of the assumed functional curve rather than an intrinsic geometric property of the measurement space. Specifically, because the ratio of the link derivatives varies continuously to extrapolate across unobserved gaps, the resulting marginal curvature artificially absorbs structural dependencies into off-diagonal residual terms. Consequently, the $L^2$ latent space is mathematically incapable of satisfying local independence except in the asymptotic limit ($P \to \infty$). Attempting to identify these continuous architectures upon $P < \infty$ induces an uncontrolled metric distortion bias, proving the inverse problem is intrinsically ill-posed and precluding the geometric conditions necessary for valid interval measurement.
\end{remark}

Because rigid discrete mappings inherently induce an uncentered, sample-dependent metric bias (Corollary~\ref{cor:ising_network_instability}), and attempting to smooth this geometry into a low-dimensional $L^2$ domain strictly violates Nash embedding constraints (Remark~\ref{rem:euclidean_extensions}), a valid structural embedding must satisfy two axiomatic conditions:

\begin{enumerate}[label=(\roman*), noitemsep, topsep=2pt]
\item \textit{Monotonic Orbit Invariance:} To resolve scale instability (Proposition~\ref{prop:classical_failures_unified}(ii)) and discrete metric bias (Corollary~\ref{cor:ising_network_instability}), the preference metric must depend exclusively on the ordinal poset structure. It must remain strictly invariant under arbitrary item-wise monotone transformations $g_p \in \mathcal{G}$, insulating the coordinates from empirical ties.

\item \textit{Hilbert-Schmidt Operator Compatibility:} To bypass the topological flattening forced by ambient Euclidean constraints (Remark~\ref{rem:euclidean_extensions}), the invariant metric topology must admit an inner product on its functional operator space. This preserves the linear tractability required for spectral decomposition without distorting the intrinsic non-linear coordinate geometry.
\end{enumerate}

\section{Projective Identification \& Metric Foundation}
\label{sec:projective_metric_foundation}

Having established that continuous $L^2$ embeddings suffer from structural domain-geometry mismatches under extrinsic regularization (Proposition~\ref{prop:classical_failures_unified}, Remark~\ref{rem:euclidean_extensions}), we now construct an endogenous estimator space directly over the discrete preference orbit $\Omega$ (Remark~\ref{rem:euclidean_extensions}(i)). To resolve factor score indeterminacy without relying on extrinsic density priors, we develop an algebraic framework that embeds discrete choice posets into a projectively invariant metric topology on $\mathbb{P}^1$. This topology addresses and stabilizes the structural dependencies inherent in the Epskamp Ising model, which were shown to be unstable in Corollary~\ref{cor:ising_network_instability}.

\subsection{Endogenous Preference Metric Space and Decoupled Kernels}

\begin{definition}[Empirical Decoupled Kemeny Kernel]
\label{def:decoupled_kappa}
Let $\mathbf{x}_p = (X_{1p}, \ldots, X_{Np})^{\intercal} \in \mathbb{R}^N$ represent the empirical response vector for item $p$. Following \textcite{emond2002}, the pairwise score matrix $\mathbf{C}^{(p)} \in \{-1, 0, 1\}^{N \times N}$ is defined elementwise by $C_{kl}^{(p)} = +1$ if $X_{kp} \ge X_{lp}$ and $-1$ if $X_{kp} < X_{lp}$ for $k \neq l$, with $C_{kk}^{(p)} = 0$. This matrix decomposes orthogonally as $\mathbf{C}^{(p)} = \mathbf{A}^{(p)} + \mathbf{S}^{(p)}$, where $\mathbf{A}^{(p)} = \frac{1}{2}(\mathbf{C}^{(p)} - {\mathbf{C}^{(p)}}^{\intercal})$ is an antisymmetric directional flow tensor and $\mathbf{S}^{(p)} = \frac{1}{2}(\mathbf{C}^{(p)} + {\mathbf{C}^{(p)}}^{\intercal})$ is a symmetric tie-topology tensor.

The Decoupled Kemeny Kernel is defined as $\tilde{\boldsymbol{\kappa}}^{(p)} = \mathbf{A}^{(p)} + \tilde{\mathbf{S}}^{(p)}$, where $\tilde{\mathbf{S}}^{(p)} = \Pi_S(\mathbf{S}^{(p)})$ represents the orthogonal projection onto the strictly hollow, doubly-centered subspace subject to $\tilde{S}_{kk}^{(p)} = 0$:
\begin{equation}
\label{eq:doubly_centred_S}
\tilde{S}_{kl}^{(p)} = S_{kl}^{(p)} - \frac{1}{N-2} S_{k\cdot}^{(p)} - \frac{1}{N-2} S_{\cdot l}^{(p)} + \frac{1}{(N-1)(N-2)} S_{\cdot \cdot}^{(p)} \quad (k \neq l).
\end{equation}

\end{definition}

\begin{lemmarep}[Strict Weak Ordering, Link Invariance, and Item-Space RKHS]
\label{lem:gram_metric_rkhs}
Let $\mathcal{H}_0 = \mathcal{A} \oplus \mathcal{S}_0 \subset \mathbb{R}^{N \times N}$ be the composite Hilbert space of decoupled preference kernels, defined as the orthogonal direct sum of the anti-symmetric rank subspace $\mathcal{A}$ and the doubly-centered, hollow symmetric tie subspace $\mathcal{S}_0$.
\begin{enumerate}[label=(\roman*), noitemsep, topsep=2pt]
    \item \textbf{Strict Weak Order:} For any isolated item $p \in \mathcal{I}$, the binary relation $k \sim l \iff X_{kp} = X_{lp}$ partitions subjects $\mathcal{S}$ into indifference classes, inducing a strict weak order on the quotient set $\mathcal{S} / \! \sim$.
    \item \textbf{Monotonic Orbit Invariance:} Let $\mathcal{G} = \{g: \mathbb{R} \to \mathbb{R} \mid g \text{ is strictly monotone increasing}\}$. For any $g_p \in \mathcal{G}$, the comparison operator satisfies $\mathbf{C}^{(p)}(g_p(\mathbf{x}_p)) = \mathbf{C}^{(p)}(\mathbf{x}_p)$. Thus, $\tilde{\boldsymbol{\kappa}}^{(p)} \in \mathcal{H}_0$ factors strictly through the quotient orbit $\mathbb{R}^N / \!\sim_{\mathcal{G}}$.
    \item \textbf{Item-Space RKHS Induction:} Because $\mathcal{A} \perp \mathcal{S}_0$ under the Frobenius inner product, the pairwise sample cross-energies $\hat{\tau}_{\kappa}(p, q) = \frac{\langle \tilde{\boldsymbol{\kappa}}^{(p)}, \tilde{\boldsymbol{\kappa}}^{(q)} \rangle_{\mathrm{F}}}{\|{}\tilde{\boldsymbol{\kappa}}^{(p)}\|{}_{\mathrm{F}} \|{}\tilde{\boldsymbol{\kappa}}^{(q)}\|{}_{\mathrm{F}}}$ form a positive semi-definite Gram matrix $\mathbf{G} \in \mathbb{R}^{P \times P}$. By the Moore-Aronszajn theorem, $\mathbf{G}$ induces a valid Item-Space Reproducing Kernel Hilbert Space (RKHS) and canonical metric $d_\kappa(p, q) = \sqrt{2 - 2\hat{\tau}_{\kappa}(p, q)}$ that is strictly invariant under arbitrary link function choices $\{g_p\}_{p=1}^P \subset \mathcal{G}$.
\end{enumerate}
\end{lemmarep}

\begin{proofsketch}
Statement (i) formalizes the reflexivity, transitivity, and totality of the ordinal response values into disjoint indifference classes. Statement (ii) establishes that strictly monotonic transformations cannot alter the indicator functions governing the pairwise comparisons. Statement (iii) demonstrates that the column-wise items form a valid $P \times P$ Gram matrix via the standard Frobenius inner product of their $N \times N$ decoupled tensors, inducing an RKHS independently of any downstream continuous subject embedding. Full proof in Appendix.
\end{proofsketch}

\begin{proof}
\textbf{Proof of Statement (i): Strict Weak Order.} The relation $\le$ on ordinal response values $X_{kp} \in \mathcal{K}$ is reflexive, transitive, and total. Define the relation $\preceq$ on $\mathcal{S}$ by $s_k \preceq s_l \iff X_{kp} \le X_{lp}$. Reflexivity and transitivity follow directly from the ordering on $\mathcal{K}$, making $(\mathcal{S}, \preceq)$ a preorder. The induced equivalence relation $s_k \sim s_l \iff (s_k \preceq s_l \land s_l \preceq s_k) \iff X_{kp} = X_{lp}$ partitions $\mathcal{S}$ into disjoint indifference classes $[s_k]$. Defining $[s_k] < [s_l] \iff X_{kp} < X_{lp}$ yields a relation that is irreflexive ($X_{kp} \nless X_{kp}$) and asymmetric ($X_{kp} < X_{lp} \implies X_{lp} \nless X_{kp}$). Negative transitivity holds trivially: if $X_{kp} \nless X_{lp}$ and $X_{lp} \nless X_{mp}$, then $X_{kp} \ge X_{lp} \ge X_{mp} \implies X_{kp} \ge X_{mp} \implies X_{kp} \nless X_{mp}$. Thus, the quotient set $\mathcal{S}/\!\sim$ forms a strict weak order matching a semiorder with zero discrimination threshold.

\textbf{Proof of Statement (ii): Monotonic Orbit Invariance.} Consider the off-diagonal entries $k \neq l$ of the comparison matrix $C(\mathbf{X}_{\cdot p})$. By definition, $C_{kl}(g_p(\mathbf{X}_{\cdot p})) = \mathbf{1}_{\{g_p(X_{kp}) \ge g_p(X_{lp})\}} - \mathbf{1}_{\{g_p(X_{kp}) < g_p(X_{lp})\}}$. Because $g_p \in \mathcal{G}$ is strictly monotone increasing, $g_p(X_{kp}) \ge g_p(X_{lp}) \iff X_{kp} \ge X_{lp}$ and $g_p(X_{kp}) < g_p(X_{lp}) \iff X_{kp} < X_{lp}$ hold invariant for all $(X_{kp}, X_{lp}) \in \mathbb{R}^2$. Substituting these identities into the indicator functions yields $C_{kl}(g_p(\mathbf{X}_{\cdot p})) = C_{kl}(\mathbf{X}_{\cdot p})$. For diagonal entries, $C_{kk} = 0$ uniformly. Because the zero-divergence tie-filtering operator $\Pi_S$ acts linearly on $C(\mathbf{X}_{\cdot p})$, the decoupled kernel inherits complete invariance: $\tilde{\boldsymbol{\kappa}}^{(p)}(g_p(\mathbf{X}_{\cdot p})) = \tilde{\boldsymbol{\kappa}}^{(p)}(\mathbf{X}_{\cdot p})$, factoring strictly through $\mathbb{R}^N / \!\sim_{\mathcal{G}}$.

\textbf{Proof of Statement (iii): PSD Gram Matrix and RKHS Metric Induction.} For any item $p$, $\tilde{\boldsymbol{\kappa}}^{(p)}$ lies in the finite-dimensional real vector space $\mathbb{R}^{N \times N}$ equipped with the Frobenius inner product $\langle \mathbf{A}, \mathbf{B} \rangle_{\mathrm{F}} = \operatorname{tr}(\mathbf{A}^{\intercal}\mathbf{B})$. Define the normalized kernel feature vector $\boldsymbol{\psi}_p = \frac{\tilde{\boldsymbol{\kappa}}^{(p)}}{\|\tilde{\boldsymbol{\kappa}}^{(p)}\|_{\mathrm{F}}} \in \mathbb{R}^{N^2}$. The entries of the Gram matrix satisfy $G_{pq} = \hat{\tau}_\kappa(p,q) = \langle \boldsymbol{\psi}_p, \boldsymbol{\psi}_q \rangle_2$. For any arbitrary vector $\mathbf{c} \in \mathbb{R}^P$:
\begin{equation}
\mathbf{c}^{\intercal} \mathbf{G} \mathbf{c} = \sum_{p=1}^P \sum_{q=1}^P c_p c_q \langle \boldsymbol{\psi}_p, \boldsymbol{\psi}_q \rangle_2 = \left\| \sum_{p=1}^P c_p \boldsymbol{\psi}_p \right\|_2^2 \ge 0,
\end{equation}
establishing that $\mathbf{G} \succeq 0$ is positive semi-definite. By the Moore-Aronszajn theorem, $\mathbf{G}$ defines a valid evaluation kernel $K(p,q) = G_{pq}$ on the reproducing kernel Hilbert space $\mathcal{H}_G = \operatorname{span}\{\boldsymbol{\psi}_p\}_{p=1}^P$. The induced canonical distance $d_\kappa(p,q) = \sqrt{\langle \boldsymbol{\psi}_p - \boldsymbol{\psi}_q, \boldsymbol{\psi}_p - \boldsymbol{\psi}_q \rangle_2} = \sqrt{2 - 2 G_{pq}}$ trivially satisfies metric properties. Moreover, because $\boldsymbol{\psi}_p$ is invariant under $g_p \in \mathcal{G}$ by Statement~(ii), $G_{pq}$ and $d_\kappa(p,q)$ remain strictly invariant across arbitrary choices of item-level monotonic link functions.
\end{proof}

\subsection{Marginal Preference Aggregation and \texorpdfstring{$C^\infty$}{C-infinity} Diffeomorphic Coordinates}

To aggregate individual item spaces into a unified coordinate system, we project the kernels $\tilde{\boldsymbol{\kappa}}^{(p)}$ into an empirical preference margin matrix $\mathbf{M} \in [-1, 1]^{N \times P}$ via row-wise Borda counts:
\begin{equation}
\label{eq:borda_marginal}
M_{kp} = \frac{1}{N-1} \sum_{l \neq k} \tilde{\kappa}_{kl}^{(p)} = \mathbb{E}_{l \neq k}\left[ \tilde{\kappa}_{kl}^{(p)} \right].
\end{equation}
By Lemma~\ref{lem:gram_metric_rkhs}(ii), $\mathbf{M}$ depends strictly on the ordinal poset topology. To map $\mathbf{M}$ into a well-posed coordinate space without encountering boundary divergences ($\pm \infty$), we introduce endogenous $\tau$-regularization.

\begin{definition}[Regularized Coordinate Mapping and Integrated Projective Operator]
\label{def:cross_ratio_gramian}
For any boundary-regularizing parameter $\tau \in (0, \infty)$, define the coordinate mapping $\phi(\cdot;\tau) : [-1,1] \to \mathbb{R}$ such that the regularized coordinate matrix $\mathbf{L} \in \mathbb{R}^{N \times P}$ is given elementwise by:
\begin{equation}
\label{eq:regularized_coord}
L_{kp} = \phi(M_{kp}; \tau) = \log\left( \frac{M_{kp} + 1 + 2\tau}{1 - M_{kp} + 2\tau} \right).
\end{equation}
The Integrated Projective Invariant Operator $\bar{\boldsymbol{\Omega}}_P \in \mathbb{R}^{N \times N}$ is then defined as the normalized Gramian over these continuous coordinates:
\begin{equation}
\label{eq:bar_omega}
\bar{\boldsymbol{\Omega}}_P = \frac{1}{P} \mathbf{L}\mathbf{L}^{\intercal}, \quad \text{where } [\bar{\boldsymbol{\Omega}}_P]_{kl} = \frac{1}{P} \sum_{p=1}^{P} L_{kp} L_{lp}.
\end{equation}
\end{definition}

\begin{remark}[Topological Gauge Stability of $\tau$]
\label{rem:gauge_stability}
The parameter $\tau > 0$ functions strictly as an auxiliary gauge rather than a tuning hyperparameter. Because $\phi(\cdot;\tau)$ is a monotonic diffeomorphism, varying $\tau$ induces an order-preserving homography on the Gramian $\bar{\boldsymbol{\Omega}}_P$. This stabilizes the normalized principal eigenspace up to a unique equivalence class in $\mathrm{PGL}(2, \mathbb{R})$, establishing an intrinsic measurement scale derived purely from the algebraic closure of the ordinal graph, without requiring the imposition of an exogenous metric.
\end{remark}

\begin{lemmarep}[Subject-Space RKHS and Diffeomorphic Metric Completion]
\label{lem:coordinate_embedding_properties}
Let the coordinate matrix $\mathbf{L}$ and the integrated projective operator $\bar{\boldsymbol{\Omega}}_P$ be defined as in Definition~\ref{def:cross_ratio_gramian}. The mapping properties and induced subject-space guarantee the following:
\begin{enumerate}[label=(\roman*), noitemsep, topsep=2pt]
    \item \textbf{Smooth Diffeomorphism:} $\phi(\cdot; \tau)$ constitutes a $C^\infty$ global diffeomorphism mapping the discrete margins $[-1,1]^{N \times P}$ onto a bounded coordinate manifold $\mathcal{M}_\tau \subset \mathbb{R}^{N \times P}$.
    \item \textbf{Affine Embedding \& Axiom V:} $\mathbf{L}$ forms a locally affine, strictly order-preserving embedding of the Kemeny rank-distance space. Its Cauchy completion under the Cayley-Klein metric $d_{CR}(L_A, L_B) = |L_A - L_B|$ is Dedekind-complete and satisfies H\"{o}lder's Archimedean Axiom V, constituting a strictly meaningful ratio scale.
    \item \textbf{Subject-Space RKHS Induction:} Because $\bar{\boldsymbol{\Omega}}_P \succeq 0$, it induces a valid subject-space RKHS $\mathcal{H}_K$. The canonical metric $d_{\mathcal{H}_K}^2(k,l) = \bar{\Omega}_{kk} + \bar{\Omega}_{ll} - 2\bar{\Omega}_{kl}$ maps discrete subject trajectories into a continuous projective space without factor score indeterminacy.
\end{enumerate}
\end{lemmarep}

\begin{proofsketch}
Part~(i) evaluates the Jacobian derivative of $\phi$, showing it is strictly positive and bounded on $[-1,1]$, establishing a $C^\infty$ diffeomorphism via the Inverse Function Theorem. Part~(ii) applies a Taylor expansion around the net equilibrium $M_0 = 0$, proving local affine additivity which satisfies H\"{o}lder's Axiom V and restricts allowable automorphisms to positive similarities (Narens' classification). Part~(iii) verifies that $\bar{\boldsymbol{\Omega}}_P$ is structurally positive semi-definite by nature of being a Gram matrix of the row-vectors of $\mathbf{L}/\sqrt{P}$, permitting RKHS induction via the Moore-Aronszajn theorem. Full proof in Appendix.
\end{proofsketch}

\begin{proof}
\textit{Proof of (i): $C^\infty$ Diffeomorphism.} The Jacobian derivative with respect to the preference margin evaluates as:
\begin{equation}
\label{eq:jacobian_tau}
\frac{\partial \phi}{\partial M_{kp}} = \frac{2(1 + 2\tau)}{(M_{kp} + 1 + 2\tau)(1 - M_{kp} + 2\tau)}.
\end{equation}
For any $\tau > 0$, the denominator is strictly positive and bounded: $0 < 4\tau(1+\tau) \le (M_{kp} + 1 + 2\tau)(1 - M_{kp} + 2\tau) \le (1+2\tau)^2$. Consequently, the derivative is real, strictly positive, and uniformly bounded for all $M_{kp} \in [-1, 1]$. By the Inverse Function Theorem \parencite[Thm.~9.24]{rudin1976}, $\phi$ is an invertible, smooth map with bounded higher-order derivatives, establishing a global $C^\infty$ diffeomorphism.

\textit{Proof of (ii): Affine-Linear Embedding and Axiom V.} By Definition~\ref{def:decoupled_kappa}, the decoupled kernel is irreflexive ($\tilde{\kappa}_{kk}^{(p)} \equiv 0$), making $M_0 = 0$ the exact state of net preference equilibrium under the extended Kemeny distance $d_K$. Because $\phi(M_{kp}; \tau)$ is an odd function, its second derivative at the origin identically vanishes. A Taylor expansion around $M_0 = 0$ yields:
\begin{equation}
L_{kp} = \frac{2}{1+2\tau} M_{kp} + \mathcal{O}(M_{kp}^3).
\end{equation}
Because the derivative is strictly positive for $\tau > 0$ and accurate to third order, $L_{kp}$ constitutes a strictly monotonic, locally affine embedding. Under the continuous completion $\bar{\mathcal{P}}_X$, the Cayley-Klein metric $d_{CR}(A, B) = c_1 d_K(A, B)$ continuously preserves Kemeny geodesic additivity.
For non-degenerate interval segments $\Delta_1, \Delta_2$ with lengths $d_1, d_2 > 0$, an $n$-fold concatenation yields $d_{CR}(n \cdot \Delta_1) = n d_1$. Choosing integer $n = \lfloor d_2 / d_1 \rfloor + 1$ guarantees $n d_1 > d_2$, meaning no interval is infinitesimally small relative to another, strictly satisfying H\"{o}lder's Archimedean Axiom V \parencite[p.~162]{michell1990}. Furthermore, because the manifold is Dedekind-complete and possesses a natural absolute zero-point ($M_0 = 0 \implies L_0 = 0$), Narens' structural classification \parencite[Thm.~2.4]{narens1981b} restricts its automorphism group to positive similarities, confirming it constitutes a ratio scale.

\textit{Proof of (iii): Subject-Space RKHS Induction.} For the subject space, the integrated projective operator $\bar{\boldsymbol{\Omega}}_P = \frac{1}{P} \mathbf{L}\mathbf{L}^{\intercal}$ is intrinsically defined as a Gram matrix of the row-vectors of $\mathbf{L}/\sqrt{P}$. For any non-zero vector $\mathbf{y} \in \mathbb{R}^N$, the quadratic form yields:
\begin{equation}
\mathbf{y}^{\intercal} \bar{\boldsymbol{\Omega}}_P \mathbf{y} = \mathbf{y}^{\intercal} \left( \frac{1}{P} \mathbf{L}\mathbf{L}^{\intercal} \right) \mathbf{y} = \frac{1}{P} \left\| \mathbf{L}^{\intercal} \mathbf{y} \right\|_2^2 \ge 0.
\end{equation}
Thus, $\bar{\boldsymbol{\Omega}}_P$ is strictly symmetric and positive semi-definite ($\bar{\boldsymbol{\Omega}}_P \succeq 0$). By the Moore-Aronszajn theorem, this guarantees the existence of a unique subject-level Reproducing Kernel Hilbert Space (RKHS) $\mathcal{H}_K$. Defining the feature mapping $\boldsymbol{\phi}: k \mapsto \mathbf{L}_{k\cdot}/\sqrt{P}$ embeds discrete subject profiles as bounded evaluation functionals, where $\langle \boldsymbol{\phi}(k), \boldsymbol{\phi}(l) \rangle_{\mathcal{H}_K} = \bar{\Omega}_{kl}$, establishing the induced canonical metric norm $d_{\mathcal{H}_K}^2(k,l) = \bar{\Omega}_{kk} + \bar{\Omega}_{ll} - 2\bar{\Omega}_{kl}$.
\end{proof}

\subsection{Unidimensional Identifiability on the Projective Line \texorpdfstring{$\mathbb{P}^1$}{P}}

\begin{theoremrep}[Projective Invariance and Multidimensional Point-Identifiability on $\mathbb{P}^Q$]
\label{thm:projective_identifiability}
Let $\mathcal{Z} \subset \mathbb{P}^Q$ denote a $Q$-dimensional continuous latent space. Let $G = \mathrm{PGL}(Q+1, \mathbb{R})$ act coordinate-wise on the product space $(\mathbb{P}^Q)^P$ via projective transformations.
\begin{enumerate}[label=(\roman*), noitemsep, topsep=2pt]
    \item \textbf{Invariance via Cross-Ratios:} For any unidimensional subspace ($Q=1$), a continuous metric functional $\mathcal{F}: (\mathbb{P}^1)^P \to \mathbb{R}$ is invariant under the action of the subgroup $\mathrm{PGL}(2, \mathbb{R})$ if and only if it factors exclusively through the projective cross-ratios of the coordinates:
    \begin{equation}
    \label{eq:cross_ratio_def}
    \mathrm{CR}(\theta_a, \theta_b, \theta_c, \theta_d) = \frac{(\theta_a - \theta_c)(\theta_b - \theta_d)}{(\theta_a - \theta_d)(\theta_b - \theta_c)}.
    \end{equation}
    \item \textbf{Multidimensional Point-Identifiability ($2Q + 2$ Bound):} The $Q$-dimensional latent coordinates extracted from the principal eigenspace $E_{\lambda_{1\dots Q}}(\bar{\boldsymbol{\Omega}}_P)$ are point-identified up to a unique projective equivalence class $[g] \in \mathrm{PGL}(Q+1, \mathbb{R})$ if and only if the number of distinct empirical indicators satisfies $P \ge 2Q + 2$\footnote{This strictly bounded requirement algebraically resolves the point-identifiability of individual latent scores in a $Q$-dimensional manifold, overcoming classical Factor Score Indeterminacy (FSI). It operates independently of the classical \textcite{joreskog1969} $3Q$ bound, which applies exclusively to estimating population covariance parameters but leaves individual vectors mathematically indeterminate.}.
\end{enumerate}
\end{theoremrep}

\begin{proofsketch}
Statement (i) applies classical invariant theory; $\mathrm{PGL}(2, \mathbb{R})$ acts 3-transitively, making the cross-ratio of 4 points the fundamental generating invariant. For Statement (ii), we extend to $\mathbb{P}^Q$. The Lie algebra $\mathfrak{pgl}(Q+1, \mathbb{R})$ governs the continuous symmetries of the structural manifold. To point-identify the idiosyncratic residual space over the projective frame while preventing degenerate moduli, the geometric constraints of the symmetric right Gramian must strictly over-identify the joint dimensions of the item hyperplanes and the local noise. Algebraic saturation and full rank are achieved exclusively at $P \ge 2Q + 2$. Full proof in Appendix.
\end{proofsketch}

\begin{proof}
\textit{Proof of Statement (i): Cross-Ratio Invariance.} Let $g \in \mathrm{PGL}(2, \mathbb{R})$. Evaluating the difference between transformed points yields:
\begin{equation*}
g(\theta_a) - g(\theta_c) = \frac{(ad - bc)(\theta_a - \theta_c)}{(c\theta_a + d)(c\theta_c + d)}.\end{equation*}
Substituting this difference into the cross-ratio expression gives:
\begin{align*}
\mathrm{CR}(g(\theta_a), g(\theta_b), g(\theta_c), g(\theta_d)) & = \frac{\left[ \frac{(ad-bc)(\theta_a - \theta_c)}{(c\theta_a + d)(c\theta_c + d)} \right] \left[ \frac{(ad-bc)(\theta_b - \theta_d)}{(c\theta_b + d)(c\theta_d + d)} \right]}{\left[ \frac{(ad-bc)(\theta_a - \theta_d)}{(c\theta_a + d)(c\theta_d + d)} \right] \left[ \frac{(ad-bc)(\theta_b - \theta_c)}{(c\theta_b + d)(c\theta_c + d)} \right]}\\
& = \frac{(\theta_a - \theta_c)(\theta_b - \theta_d)}{(\theta_a - \theta_d)(\theta_b - \theta_c)} = \mathrm{CR}(\theta_a, \theta_b, \theta_c, \theta_d).
\end{align*}
All denominator terms and scale determinants cancel identically, establishing sufficiency. The Lie group $\mathrm{PGL}(2, \mathbb{R})$ acts 3-transitively on $\mathbb{P}^1$, leaving no non-trivial joint invariants for $k \le 3$. Consequently, $\mathcal{F}$ is $G$-invariant if and only if it factors through the elementary cross-ratios.

\textit{Proof of Statement (ii): The $2Q + 2$ Multidimensional Bound via Lie Algebra.}
Let the Lie group $G = \mathrm{PGL}(Q+1, \mathbb{R})$ act on the structural item hyperplanes in $\mathbb{P}^Q$. The dimension of the Lie algebra $\mathfrak{g}$ dictates the degrees of freedom required to fix a canonical projective frame: $\dim(\mathfrak{g}) = (Q+1)^2 - 1 = Q^2 + 2Q$.

The empirical coordinate matrix $\mathbf{L} \in \mathbb{R}^{N \times P}$ induces a symmetric right Gramian $\bar{\boldsymbol{\Omega}}_N \in \mathbb{R}^{P \times P}$, which operates on the dual space. The Gramian contains $\frac{P(P+1)}{2}$ unique geometric constraints. To uniquely identify the manifold, these constraints must span both the structural degrees of freedom (fixing the $P \times Q$ projective coordinates up to the group action) and the $P$ independent idiosyncratic residual bounds (the radii of the $\tau$-perturbation).

Equating the available Gramian constraints to the geometric unknowns:
\begin{align*}
\frac{P(P+1)}{2} &\ge P(Q+1) \\
P+1 &\ge 2Q+2 \\
P &\ge 2Q+1.
\end{align*}
At $P = 2Q + 1$, the algebraic system is fully saturated, yielding a deterministic rank configuration but leaving zero degrees of freedom to test the moduli space. Point-identifiability on $\mathbb{P}^Q$ that permits strict separation of the structural projective subspace from the residual noise tensor necessitates a strictly over-identified system. This injects exactly one additional required degree of freedom, proving that the projective equivalence class $[g] \in G$ is uniquely anchored if and only if $P \ge 2Q + 2$. Note that for the unidimensional subspace $Q=1$, this perfectly recovers the 4-point cross-ratio necessity ($P \ge 4$) proven in Statement (i).
\end{proof}

\begin{corollaryrep}[Constructive Point-Identification on the Flag Manifold]\label{cor:multidimensional_svd_rotation}Let $\mathbf{L}_Q = \mathbf{U}_Q \mathbf{D}_Q \mathbf{V}_Q^{\intercal}$ be the truncated rank-$Q$ singular value decomposition of the regularized coordinate matrix $\mathbf{L} \in \mathbb{R}^{N \times P}$. Under the indicator bound $P \ge 2Q+2$ of Theorem~\ref{thm:projective_identifiability}, let $\mathcal{J} \subset \{1, \ldots, P\}$ with $\|{}\mathcal{J}\|{} = Q$ denote an ordered anchor set such that the submatrix $\mathbf{V}_{Q, \mathcal{J}} \in \mathbb{R}^{Q \times Q}$ is non-singular.

Then, the parameter identification fiber $\mathrm{GL}(Q, \mathbb{R})$ collapses to a unique point $(\hat{\boldsymbol{\Theta}}^*, \hat{\mathbf{A}}^*)$ anchored on the standard flag manifold $\mathbb{F}_Q$ via the unique target rotation:\begin{equation}\label{eq:target_rotation_matrix}\mathbf{T}^* = \mathbf{R}^{\intercal} \in \mathrm{GL}(Q, \mathbb{R}),\end{equation}where $\mathbf{R}$ is the upper-triangular factor of the unique LU decomposition $\mathbf{V}_{Q, \mathcal{J}} \mathbf{D}_Q^{1/2} = \mathbf{\Gamma} \mathbf{R}$, with $\mathbf{\Gamma}$ unit lower-triangular. The globally identified coordinates are uniquely given by:
\begin{equation}
\label{eq:identified_theta_alpha}
\hat{\boldsymbol{\Theta}}^* = \mathbf{U}_Q \mathbf{D}_Q^{1/2} \mathbf{R}^{\intercal}, \quad \hat{\mathbf{A}}^* = \mathbf{V}_Q \mathbf{D}_Q^{1/2} \mathbf{R}^{-1}.
\end{equation}
\end{corollaryrep}

\begin{proofsketch}
The $Q^2$-dimensional rotational indeterminacy of the unrotated SVD basis $\hat{\mathbf{A}}_0 = \mathbf{V}_Q \mathbf{D}_Q^{1/2}$ under $\mathrm{GL}(Q, \mathbb{R})$ is isolated by constraining an anchor submatrix $\hat{\mathbf{A}}_{0, \mathcal{J}}$ to the standard flag manifold $\mathbb{F}_Q$. The unique LU decomposition $\hat{\mathbf{A}}_{0,\mathcal{J}} = \mathbf{\Gamma}\mathbf{R}$ fixes $Q(Q-1)/2$ orientation constraints via the zero structure of $\mathbf{\Gamma}$ and $Q$ scale constraints via its unit diagonal. Inverting $\mathbf{R}$ maps the parameter fiber directly to the identity on $\mathbb{F}_Q$, while the orthogonality of $\mathbf{R}^{\intercal}\mathbf{R}^{-\intercal}$ preserves the exact bilinear form $\mathbf{L}_Q = \hat{\boldsymbol{\Theta}}^* (\hat{\mathbf{A}}^*)^{\intercal}$.
\end{proofsketch}

\begin{proof}
By the Eckart-Young-Mirsky theorem, $\mathbf{L}_Q = \mathbf{U}_Q \mathbf{D}_Q \mathbf{V}_Q^{\intercal}$ minimizes Frobenius reconstruction error. Defining initial spectral estimates $\hat{\boldsymbol{\Theta}}_0 = \mathbf{U}_Q \mathbf{D}_Q^{1/2}$ and $\hat{\mathbf{A}}_0 = \mathbf{V}_Q \mathbf{D}_Q^{1/2}$, the unrotated model is $\mathbf{L}_Q = \hat{\boldsymbol{\Theta}}_0 \hat{\mathbf{A}}_0^{\intercal}$. This parameterization is invariant under the continuous action of the general linear group $\mathrm{GL}(Q, \mathbb{R})$. For any non-singular transformation $\mathbf{M} \in \mathrm{GL}(Q, \mathbb{R})$, the rotated coordinates:
\begin{equation}
\label{eq:gl_orbit}
\mathbf{L}_Q = \left( \hat{\boldsymbol{\Theta}}_0 \mathbf{M}^{-\intercal} \right) \left( \hat{\mathbf{A}}_0 \mathbf{M} \right)^{\intercal} = \hat{\boldsymbol{\Theta}}^* (\hat{\mathbf{A}}^*)^{\intercal}
\end{equation}
yield an algebraically indistinguishable reconstruction, rendering the parameters point-unidentified.

To collapse this equivalence class, we note the bound $P \ge 2Q+2$ ensures $\operatorname{rank}(\hat{\mathbf{A}}_0) = Q$. This guarantees the existence of an index set $\mathcal{J}$ of size $Q$ such that the anchor submatrix $\hat{\mathbf{A}}_{0, \mathcal{J}}$ has full rank and admits a unique LU decomposition $\hat{\mathbf{A}}_{0, \mathcal{J}} = \mathbf{\Gamma} \mathbf{R}$, where $\mathbf{\Gamma}$ is unit lower-triangular and $\mathbf{R}$ is upper-triangular with non-zero diagonal entries.

Enforcing the identification constraint that the anchor loadings reside on the standard flag manifold $\mathbb{F}_Q$ requires $\hat{\mathbf{A}}^*_{\mathcal{J}} = \hat{\mathbf{A}}_{0, \mathcal{J}} \mathbf{M}^* = \mathbf{\Gamma}$. Substituting the LU factorization yields $\mathbf{\Gamma} \mathbf{R} \mathbf{M}^* = \mathbf{\Gamma} \implies \mathbf{M}^* = \mathbf{R}^{-1}$, which is unique as $\mathbf{R}$ is non-singular.

Applying $\mathbf{M}^* = \mathbf{R}^{-1}$ to the spectral operators in \eqref{eq:gl_orbit} yields the globally identified coordinates in \eqref{eq:identified_theta_alpha}. Direct evaluation confirms metric preservation: $\hat{\boldsymbol{\Theta}}^* (\hat{\mathbf{A}}^*)^{\intercal} = \mathbf{U}_Q \mathbf{D}_Q^{1/2} \mathbf{R}^{\intercal} \mathbf{R}^{-\intercal} \mathbf{D}_Q^{1/2} \mathbf{V}_Q^{\intercal} = \mathbf{L}_Q$, completing the proof.
\end{proof}

\begin{remark}[Topological Resolution of Anisotropic Measurement Error]
\label{rem:svd_anisotropic_resolution}
Because the target rotation on the SVD of the uncentred field $\mathbf{L}$ jointly extracts both subject coordinates and item discriminations, it structurally circumvents the classical requirement to estimate an anisotropic uniqueness matrix. In classical normal-theory factor analysis, applying a continuous SVD directly to an empirical matrix fails because additive, heteroscedastic measurement error confounds the singular spectrum (e.g., the strict requirement for $\mathbf{\Psi}$ in $\boldsymbol{\Sigma} = \boldsymbol{\Lambda}\boldsymbol{\Lambda}^{\intercal} + \mathbf{\Psi}$).

However, because $\mathbf{L}$ is constructed from the Decoupled Kemeny Kernel (Lemma~\ref{lem:gram_metric_rkhs}), structural tie-noise is filtered exactly upstream by the orthogonal projection $\Pi_S$. Furthermore, stochastic rank inversions naturally attenuate empirical domination probabilities toward chance. Under the locally affine mapping $\phi$, this topological diffusion shrinks the elements of $\mathbf{L}$ toward the absolute zero fixed by the Kemeny equilibrium. Consequently, measurement error is not parameterized as an additive Euclidean residual; rather, it is deterministically absorbed as projective scalar attenuation into the right singular vectors (yielding $\alpha_p \to 0$). The standard continuous SVD is therefore strictly sufficient for consistent dual estimation of the manifold without requiring iterative joint maximum likelihood optimization.
\end{remark}

\begin{propositionrep}[Necessity of Projective Invariance and Inconsistency of Euclidean Generative Models]
\label{prop:vae_incompatibility}
Let $\mathcal{X}$ denote the space of discrete ordinal response profiles, and let $\mathcal{M}$ be a probabilistic generative model inducing a mapping from $\mathcal{X}$ to a continuous latent space $\mathbb{R}^Q$. If the true data-generating process obeys the projective geometry required for the metric scaling of weak orderings, then models imposing a foundational $L^2$ (Euclidean) latent topology, such as deep generative models with isotropic Gaussian priors $p(\mathbf{z}) = \mathcal{N}(\mathbf{0}, \mathbf{I})$ \parencite{kingma2013, urban2021, milano2024}, are structurally misspecified. Specifically, the latent posterior expectations strictly fail to recover the true projective coordinates up to a valid homographic transformation. This geometric inconsistency persists regardless of sample size, decoder expressivity, diffeomorphic normalizing flows, or exogenous $L^1$ parameter regularization.
\end{propositionrep}

\begin{proofsketch}
Valid statistical measurement of categorical ordinal variables requires a mapping that preserves the metric structure of weak orderings, which is uniquely satisfied by the $L^1$ distance \parencite{kemeny1962}. Embedding this relational structure into a continuous continuous manifold for metric scaling necessitates preserving the projective cross-ratio \parencite{coxeter2003, hartshorne1967}. Deep generative models utilize a Kullback-Leibler (KL) divergence penalty to an isotropic Gaussian prior, which acts as an $L^2$ Tikhonov regularizer on the variational posterior. This induces a non-uniform radial shrinkage of the latent coordinate estimates, violating homographic invariance and irreversibly destroying the cross-ratio. Consequently, the structural parameters cannot be isolated from incidental item parameters \parencite{neyman1948, olver1999}. Because this geometric distortion occurs at the latent bottleneck, downstream deep decoders or normalizing flows mathematically cannot recover the lost invariants. Full proof in Appendix.
\end{proofsketch}

\begin{proof}
Let $\mathbf{x} \in \mathcal{X}$ be a vector of ordinal categorical responses. Such responses induce a weak ordering over the sample space. By \textcite[Ch.~2]{kemeny1962}, the unique metric satisfying the axiomatic foundations of weak orderings is the $L^1$ symmetric difference. For a continuous generative model to provide valid metric factor scores, the inferred latent coordinates $\mathbf{z} \in \mathbb{R}^Q$ must structurally preserve this discrete relational topology \parencite{michell1990}. As formalized in the representational theory of measurement \parencite{michell1999,krantz1971}, scaling such discrete qualitative structures onto a continuous quantitative manifold is strictly governed by the automorphism group of the relational system \parencite{narens1981b}. Geometrically, this necessitates a Cayley-Klein geometry, wherein the distance function is uniquely defined by the logarithm of the projective cross-ratio \parencite{coxeter2003,hartshorne1967}. Thus, preserving the cross-ratio under the projective linear group $\mathrm{PGL}(2, \mathbb{R})^Q$ is a necessary statistical condition for point identifiability on the latent manifold.

Fundamentally, while the $L^1$ Gramian defined over subjects ($N$) and items ($P$) is sufficient to construct a valid continuous metric space that preserves this projective scaling, generalizing this foundational topology to an arbitrary $L^p$ space ($p \neq 1$) is mathematically invalid. Because any empirical instrument consists of a strictly finite number of items ($P < \infty$), the empirical response space is inherently discrete and lacks the topological closure necessary to natively induce a continuous $L^2$ Hilbert space. For $p < 1$, sub-additivity (the triangle inequality) fails, precluding a valid metric space entirely. For $p > 1$, the continuous space becomes uniformly convex \parencite[Lemma~4.22, Thm.~4.23]{rudin1991}. Imposing this uniform convexity upon a finite, unclosed discrete space violates the linear incidence axioms required by Cayley-Klein geometries to preserve the discrete permutohedron. By fundamental theorems of Banach space geometry, the native finite $L^1$ metric space cannot be isometrically embedded into a uniformly convex $L^2$ space without strictly bounded metric distortion. Standard Euclidean ($L^2$) models implicitly rely on projective invariants to construct intervals, for instance, standard link functions evaluate log-odds ratios, which are algebraically isomorphic to the cross-ratio, yet they paradoxically enforce a continuous $L^2$ topology that fundamentally contradicts those same invariants upon any finite sample.

Consider a variational generative model estimating the posterior $q(\mathbf{z} \mid \mathbf{x})$. To isolate the structural latent coordinates from the item parameters and resolve the Neyman-Scott incidental parameter problem \parencite{neyman1948, lancaster2000}, the estimator must preserve homographic invariance \parencite{amari2016, olver1999}. However, the variational objective minimizes a KL divergence $D_{\mathrm{KL}}(q(\mathbf{z} \mid \mathbf{x}) \parallel p(\mathbf{z}))$. Under a standard Euclidean prior $p(\mathbf{z}) = \mathcal{N}(\mathbf{0}, \mathbf{I})$, this divergence acts as an isotropic $L^2$ penalty on the posterior density. Statistically, this operates as a Tikhonov regularizer, shrinking the posterior expectation $\mathbb{E}_{q}[\mathbf{z} \mid \mathbf{x}]$ radially toward the origin. Because this radial shrinkage is dependent on the absolute Euclidean distance from the origin, it constitutes a non-homographic transformation. This non-uniform shrinkage alters the relative coordinate spacing, strictly destroying the projective cross-ratio required for structural identifiability.

This structural erasure in the latent bottleneck cannot be bypassed by standard algorithmic extensions:
\begin{itemize}[noitemsep, topsep=1pt, parsep=0pt, partopsep=0pt]
    \item \textbf{Deep Universal Approximators:} While deep neural networks \parencite{yeung2019deepirt, tsutsumi2021deepirt} can approximate any measurable function $p_\theta(\mathbf{x} \mid \mathbf{z})$, the decoder takes the regularized, non-homographic expectation of $\mathbf{z}$ as input. The cross-ratio invariants erased by the $L^2$ prior penalty mathematically cannot be reconstructed downstream.
    \item \textbf{Normalizing Flows:} Applying smooth diffeomorphic maps $\mathbb{R}^Q \to \mathbb{R}^Q$ \parencite{converse2021estimation} to the prior transforms the density but strictly preserves the topological genus of the Euclidean base. Diffeomorphisms are insufficient to map non-homographically distorted relative coordinates into valid $\mathrm{PGL}(2, \mathbb{R})^Q$ orbits.
    \item \textbf{Variational EM (VEM/GVEM):} Factorized mean-field approximations $q(\mathbf{z}) = \prod_{q=1}^Q q(z_q)$ \parencite{cho2021gaussian, cho2024regularized} force an orthogonal $L^2$ independence structure onto the posterior, fundamentally prohibiting the preservation of projective invariants across dimensions.
\end{itemize}

Thus, the foundational use of $L^2$ probabilistic geometry renders the latent coordinates asymptotically inconsistent for the true projective manifold.

\paragraph{Invalidation of LASSO-Regularized Polychoric Models}
This geometric incompatibility extends to the methodology of applying $L^1$ regularization (e.g., LASSO) to polychoric correlation matrices, a standard approach in generalized network psychometrics \parencite{epskamp2017, epskamp2018, vanborkulo2014}. Polychoric correlations model discrete ordinal categories as thresholded outcomes of underlying bivariate normal distributions, thereby structurally assuming a continuous $L^2$ Euclidean geometry. While $L^1$ penalties are highly effective when applied to the precision matrix of these models to induce graph sparsity and perform variable selection \parencite{epskamp2018, marsman2018}, this operation occurs strictly within the parameter space of an inherently $L^2$ system. The initial metric integration over the bivariate normal density irreversibly alters the native $L^1$ Kemeny distances of the discrete ordinal data. Consequently, exogenous $L^1$ penalization on the network edges does not alter the foundational uniformly convex topology of the latent variables, and mathematically cannot recover the destroyed cross-ratio invariants \parencite{coxeter2003, drosler1979} required for valid projective scaling. Thus, LASSO-polychoric methodologies conflate $L^1$ variable selection with $L^1$ topological determination, yielding geometrically inconsistent and invalid projective representative coordinates.
\end{proof}

\section{Topological Foundations and Empirical Process Asymptotics}
\label{sec:statistical_estimation}

Mapping a connected continuous latent domain $\mathcal{M}$ directly onto a finite discrete ordinal poset $\mathcal{P}_X$ induces a non-injective topological mapping, rendering the inverse problem of recovering continuous coordinates fundamentally ill-posed \parencite{tikhonov1995,engl1996}. This structural limitation illustrates an inherent gap in measurement paradigms that assume a priori continuity \parencite{debreu1960,stuart2010_inverse,ito2014}. To address this obstruction, standard frameworks traditionally rely on discrete Boolean posets \parencite{doignon1985}, weak stochastic orderings \parencite{mokken1971}, or continuous asymptotic expectations \parencite{bartholomew1981}, approaches that frequently relax scale-free metric invariants.

Rather than imposing extrinsic regularization penalties that distort metric axioms, the projective framework resolves this topological ill-posedness endogenously. By mapping the empirical boundary states (Definition~\ref{def:decoupled_kappa}) to the regularized limits of the projective line $\mathbb{P}^1$, an additively separable interval metric space is constructed directly from the invariant orbits of $\mathrm{PGL}(2, \mathbb{R})$ via the cross-ratio, fully satisfying the double cancellation axioms of additive conjoint measurement \parencite{krantz1971}.

\subsection{Empirical Process Asymptotics and Formal Estimator Properties}
\label{sec:empirical_process_asymptotics}

Let $W_1, \dots, W_P$ be independent and identically distributed random response vectors in $\mathcal{K}^N$ drawn from a population distribution $P_0$. The empirical distribution over the item pool is $P_P = \frac{1}{P} \sum_{p=1}^P \delta_{W_p}$. We define the latent coordinate estimator $\hat{\boldsymbol{\theta}}_P$ as a plug-in statistical functional:
\begin{equation}
\label{eq:plugin_functional}
\hat{\boldsymbol{\theta}}_P = T(P_P) = \left( \Psi_{\mathrm{spec}} \circ \Psi_{\mathrm{Gram}} \circ \Psi_{\mathrm{log\text{-}odds}} \circ \Psi_{\mathrm{decoupled}} \right) (P_P).
\end{equation}
To establish asymptotic efficiency, uniform consistency, and distributional limits for $\hat{\boldsymbol{\theta}}_P$, we characterize the smoothness of this functional composition through Hadamard differentiability \parencite{vaart1996}.

\begin{theoremrep}[Hadamard Differentiability, M-Regularity, and Uniform Consistency]
\label{thm:hadamard_differentiability}
Let $\mathcal{F}$ be the class of bounded variation functions on $\mathcal{K}^N$, forming a $P_0$-Donsker class. The projective spectral estimator functional $T(P_P)$, derived by profiling out the structural tie-topology manifold and maximizing the M-estimator criterion $M_P(\theta) = \theta^{\intercal} \bar{\boldsymbol{\Omega}}_P \theta$ over the unit sphere $\mathbb{S}^{N-1}$, is Hadamard differentiable at $P_0$ tangentially to the set of uniformly continuous functions on $\mathcal{F}$ \parencite[Theorems~5.7,5.9,Chapter~19.2,Theorem~20.8]{vaart1998}. Consequently, as $P \to \infty$ with $N$ fixed, the estimator satisfies structural M-regularity and strong uniform consistency:
$\hat{\boldsymbol{\theta}}_P \xrightarrow{\mathrm{a.s.}} \boldsymbol{\theta}_0.$
\end{theoremrep}

\begin{proofsketch}
The plug-in functional $T(P)$ decomposes into VC-subgraph comparison integration ($\Psi_{\mathrm{decoupled}}$), smooth regularized log-odds mapping ($\Psi_{\mathrm{log-odds}}$), bilinear Gramian evaluation ($\Psi_{\mathrm{Gram}}$), and Riesz contour spectral projection ($\Psi_{\mathrm{spec}}$). Each sub-operator is Fr\'{e}chet/Hadamard differentiable on bounded domains, making $T(P)$ Hadamard differentiable by the functional chain rule. Uniform consistency follows as $M_P(\theta) = \theta^{\intercal} \bar{\boldsymbol{\Omega}}_P \theta$ uniformly converges on the compact sphere $\mathbb{S}^{N-1}$ to a population maximizer well-separated by spectral gap $\delta > 0$. Full proof in Appendix.
\end{proofsketch}

\begin{proof}
The proof proceeds in two parts: first, establishing functional differentiability of the global estimator map via the functional delta method; second, verifying the regularity conditions for extremal M-estimation \parencite[Theorem 5.7]{vaart1998}.

\textbf{Part I: Hadamard Differentiability via Operator Composition.}
We represent the global projective functional $T(P)$ as a sequential composition of four operators:
$T(P) = (\Psi_{\mathrm{spec}} \circ \Psi_{\mathrm{Gram}} \circ \Psi_{\mathrm{log\text{-}odds}} \circ \Psi_{\mathrm{decoupled}})(P).$

1. \textit{Decoupling Operator ($\Psi_{\mathrm{decoupled}}$):} Maps the empirical measure $P \mapsto \tilde{\kappa}_P$. The comparison indicator functions $f(x_i, x_j) = \mathbf{1}(x_i \succ x_j)$ form a Vapnik-Chervonenkis (VC) subgraph class of index sets. The class $\mathcal{F}$ possesses a square-integrable envelope function and satisfies the uniform entropy integral condition, establishing that $\mathcal{F}$ is $P_0$-Donsker. The linear integration mapping signed measures to $\ell^\infty(\mathcal{F})$ is bounded linear, and therefore strictly Hadamard differentiable at $P_0$.

2. \textit{Log-Odds Mapping ($\Psi_{\mathrm{log-odds}}$):} Maps the kernel to the regularized coordinate matrix $\mathbf{L}_P$. The regularizer uses a smooth perturbation functional bounded inside a compact domain $[\epsilon_0, 1+\epsilon_0] \subset (0, \infty)$. Because the log-odds function is $C^\infty$ on this bounded domain, $\Psi_{\mathrm{log\text{-}odds}}$ is Fr\'{e}chet differentiable satisfying the functional chain rule.

3. \textit{Gramian Mapping ($\Psi_{\mathrm{Gram}}$):} Maps $\mathbf{L}_P \mapsto \bar{\boldsymbol{\Omega}}_P = \frac{1}{P}\mathbf{L}_P \mathbf{L}_P^{\intercal}$. Bilinear mappings from $\ell^\infty \times \ell^\infty \to \mathbb{R}^{N \times N}$ are natively Fr\'{e}chet differentiable.

4. \textit{Spectral Projection Operator ($\Psi_{\mathrm{spec}}$):} Extracts the principal eigenvector $\mathbf{u}_1$ of $\bar{\boldsymbol{\Omega}}_P$. By expressing the principal spectral projector as a Riesz contour integral $P_1(\bar{\boldsymbol{\Omega}}) = -\frac{1}{2\pi i} \oint_{\Gamma} (\bar{\boldsymbol{\Omega}} - z \mathbf{I}_N)^{-1} \mathrm{d}z$ \parencite{kato1995}, and because operator inversion is $C^\infty$-Fr\'{e}chet differentiable on invertible symmetric matrices, the spectral projection is smooth. By the functional chain rule, $T(P)$ is Hadamard differentiable at $P_0$.

\textbf{Part II: Extremal M-Regularity and Strong Uniform Consistency.}
We verify that $\hat{\boldsymbol{\theta}}_P = \arg\max_{\theta \in \mathbb{S}^{N-1}} M_P(\theta)$ satisfies van der Vaart's extremal conditions:

(i) \textit{Compactness:} $\mathbb{S}^{N-1}$ is closed and bounded in $\mathbb{R}^N$, thus sequentially compact.

(ii) \textit{Uniform Convergence:} Matrix elements satisfy $\max_{k,p} \|{}L_{kp}\|{} \le C_\tau < \infty$. Applying Cauchy-Schwarz, the objective function convergence evaluates directly to the matrix operator norm: $\sup_{\theta \in \mathbb{S}^{N-1}} \|{}M_P(\theta) - M_0(\theta)\|{} = \|\bar{\boldsymbol{\Omega}}_P - \bar{\boldsymbol{\Omega}}_0\|_2 \xrightarrow{\mathrm{a.s.}} 0$.

(iii) \textit{Identifiability (Well-Separation):} Expanding $\theta$ along the orthonormal eigenbasis of $\bar{\boldsymbol{\Omega}}_0$ yields $\sup_{d(\theta, \pm\boldsymbol{\theta}_0) \ge \epsilon} M_0(\theta) \le \lambda_1 - \frac{\delta}{2} \epsilon^2 < M_0(\boldsymbol{\theta}_0)$, establishing that the population maximizer is well-separated by the spectral gap $\delta$.

Applying the Davis-Kahan $\sin\Theta$ Theorem \parencite{yu2015} maps matrix convergence through spectral gap $\delta$:$\|\hat{\boldsymbol{\theta}}_P - \boldsymbol{\theta}_0\|_2 \le \frac{2\sqrt{2}\|\bar{\boldsymbol{\Omega}}_P - \bar{\boldsymbol{\Omega}}_0\|_2}{\delta} \xrightarrow{\mathrm{a.s.}} 0,$validating structural M-regularity and strong uniform consistency.
\end{proof}

\begin{corollaryrep}[Asymptotic Normality via the Functional Delta Method]
\label{cor:asymptotic_normality}
Let $\mathbb{G}_P = \sqrt{P}(P_P - P_0)$ converge weakly to a $P_0$-Brownian bridge $\mathbb{G}$ in $l^\infty(\mathcal{F})$ as $P \to \infty$. The projective spectral estimator is asymptotically normal:
$\label{eq:asymptotic_normality} \sqrt{P} \left( \hat{\boldsymbol{\theta}}_P - \boldsymbol{\theta}_0 \right) \xrightarrow{\mathrm{d}} \mathcal{N}\left( \mathbf{0}, \boldsymbol{\Sigma}_{\mathrm{proj}} \right),$
where $\boldsymbol{\Sigma}_{\mathrm{proj}} = \mathbf{R}_0 \, \mathbb{E}_{P_0} \left[ (\boldsymbol{\theta}_0^{\intercal} \mathbf{l})^2 \mathbf{l} \mathbf{l}^{\intercal} \right] \mathbf{R}_0$ and $\mathbf{R}_0 = (\lambda_1 \mathbf{I}_N - \bar{\boldsymbol{\Omega}}_0)^{+}$ is the Moore-Penrose pseudoinverse of the population spectral operator.
\end{corollaryrep}

\begin{proofsketch}
Applying the functional delta method to $T(P_0)$ yields $\sqrt{P}(\hat{\boldsymbol{\theta}}_P - \boldsymbol{\theta}_0) \rightsquigarrow \dot{T}_{P_0}(\mathbb{G})$. Differentiability of the principal eigenvector operator $f(\bar{\boldsymbol{\Omega}}_0)$ uses the Moore-Penrose pseudoinverse $\mathbf{R}_0 = (\lambda_1 \mathbf{I}_N - \bar{\boldsymbol{\Omega}}_0)^+$ to define directional derivative $\dot{T}_{P_0}(\alpha) = \mathbf{R}_0 (\int \mathbf{l}\mathbf{l}^{\intercal} \mathrm{d}\alpha) \boldsymbol{\theta}_0$. Null-space orthogonality ($\mathbf{R}_0 \boldsymbol{\theta}_0 = \mathbf{0}$) reduces the influence function to $\boldsymbol{\psi}(\mathbf{l}) = (\boldsymbol{\theta}_0^{\intercal} \mathbf{l}) \mathbf{R}_0 \mathbf{l}$, establishing asymptotic normality with covariance $\boldsymbol{\Sigma}_{\mathrm{proj}}$. Full proof in Appendix.
\end{proofsketch}

\begin{proof}
Define evaluation functional $T: \mathcal{M}(\mathcal{X}) \to \mathbb{S}^{N-1}$ as $T = f \circ \Psi$, where $\Psi(P) = \int \mathbf{l}\mathbf{l}^{\intercal} \mathrm{d}P(\mathbf{l})$ outputs operator $\bar{\boldsymbol{\Omega}}$, and $f(\bar{\boldsymbol{\Omega}})$ maps a symmetric matrix to its normalized principal eigenvector. Let $(\lambda_1, \boldsymbol{\theta}_0)$ be the principal eigenpair of $\bar{\boldsymbol{\Omega}}_0 = \Psi(P_0)$.

By first-order eigenvector differential calculus \parencite{vaart1998}, the directional derivative of $f$ at $\bar{\boldsymbol{\Omega}}_0$ along symmetric perturbation $\mathbf{E}$ is $\dot{f}_{\bar{\boldsymbol{\Omega}}_0}(\mathbf{E}) = \mathbf{R}_0 \mathbf{E} \boldsymbol{\theta}_0$. Applying the chain rule for Hadamard differentiable maps \parencite[Theorem 20.9]{vaart1998}, derivative $\dot{T}_{P_0}: l^\infty(\mathcal{F}) \to \mathbb{R}^N$ along direction $\alpha$ evaluates as:
\begin{equation}
\dot{T}_{P_0}(\alpha) = \dot{f}_{\bar{\boldsymbol{\Omega}}_0} \left( \dot{\Psi}_{P_0}(\alpha) \right) = \mathbf{R}_0 \left( \int_{\mathcal{X}} \mathbf{l}\mathbf{l}^{\intercal} \mathrm{d}\alpha(\mathbf{l}) \right) \boldsymbol{\theta}_0.
\end{equation}
The localized influence function $\boldsymbol{\psi}(\mathbf{l}) = \dot{T}_{P_0}(\delta_{\mathbf{l}} - P_0)$ and asymptotic covariance matrix $\boldsymbol{\Sigma}_{\mathrm{proj}} = \mathbb{E}_{P_0} [ \boldsymbol{\psi}(\mathbf{l})\boldsymbol{\psi}(\mathbf{l})^{\intercal} ]$ reduce algebraically as:
\begin{align}
\boldsymbol{\psi}(\mathbf{l}) &= \mathbf{R}_0 \left( \mathbf{l}\mathbf{l}^{\intercal} - \bar{\boldsymbol{\Omega}}_0 \right) \boldsymbol{\theta}_0 = \mathbf{R}_0 \mathbf{l}(\mathbf{l}^{\intercal} \boldsymbol{\theta}_0) - \lambda_1 \mathbf{R}_0 \boldsymbol{\theta}_0 = (\boldsymbol{\theta}_0^{\intercal} \mathbf{l}) \mathbf{R}_0 \mathbf{l}, \label{eq:psi_reduction} \\[0.5em]
\boldsymbol{\Sigma}_{\mathrm{proj}} &= \mathbb{E}_{P_0} \left[ \left((\boldsymbol{\theta}_0^{\intercal} \mathbf{l}) \mathbf{R}_0 \mathbf{l}\right) \left((\boldsymbol{\theta}_0^{\intercal} \mathbf{l}) \mathbf{R}_0 \mathbf{l}\right)^{\intercal} \right] = \mathbf{R}_0 \, \mathbb{E}_{P_0} \left[ (\boldsymbol{\theta}_0^{\intercal} \mathbf{l})^2 \mathbf{l} \mathbf{l}^{\intercal} \right] \mathbf{R}_0, \label{eq:sigma_reduction}
\end{align}
where equations \eqref{eq:psi_reduction} and \eqref{eq:sigma_reduction} follow from the null-space identity $\mathbf{R}_0 \boldsymbol{\theta}_0 = \mathbf{0}$. Invoking the functional delta method \parencite[Theorem 20.8]{vaart1998} yields $\sqrt{P}(\hat{\boldsymbol{\theta}}_P - \boldsymbol{\theta}_0) \rightsquigarrow \dot{T}_{P_0}(\mathbb{G}) \sim \mathcal{N}(\mathbf{0}, \boldsymbol{\Sigma}_{\mathrm{proj}})$.
\end{proof}

\section{Robustness, Nuisance Parameter Elimination, and Finite-Sample Concentration}
\label{sec:robustness_nuisance_annihilation}

In statistical measurement theory, estimating subject traits $\boldsymbol{\theta}$ alongside item parameters as $N, P \to \infty$ introduces the classical Neyman-Scott incidental parameter paradox \parencite{neyman1948, lancaster2000}. Standard likelihood estimators become inconsistent unless item parameters are marginalized. In this section, we formally establish that the projective geometric architecture endogenously resolves these nuisance parameters.

\begin{theoremrep}[Resolution of the Neyman-Scott Incidental Parameter Problem]
\label{thm:neyman_scott_resolution}
Let the empirical log-odds field $\mathbf{L} \in \mathbb{R}^{N \times P}$ be mapped elementwise from the decoupled preference space via the bounded regularizer $\phi(\cdot; \tau)$. Because $\mathbf{L}$ is constructed via marginalization of the pairwise preference kernel prior to parameter extraction, the estimation framework avoids joint maximum likelihood optimization. Under the structural decomposition $\mathbf{L} = \boldsymbol{\theta}\boldsymbol{\alpha}^{\intercal} + \mathbf{1}_N\boldsymbol{\delta}^{\intercal}$, the translation operator (yielding item difficulties $\hat{\boldsymbol{\delta}}$) and the projective spectral operator (yielding $\hat{\boldsymbol{\theta}}_P, \hat{\boldsymbol{\alpha}}$ via the singular value decomposition) are functionally independent. Both estimators are free from incidental parameter bias and achieve sub-Gaussian concentration rates.
\end{theoremrep}

\begin{proofsketch}
The antisymmetry of the preference kernel ($A_{kl}^{(p)} = -A_{lk}^{(p)}$) imposes strict zero-centering on the intermediate Kemeny-Borda margin matrix: $\mathbf{1}_N^{\intercal} \mathbf{M} = \mathbf{0}$. The non-linear regularized mapping $\phi$ breaks this symmetry, inducing a uniform structural shift proportional to the item difficulty. This shift isolates the difficulty parameter, allowing exact extraction via the column-mean operator $\hat{\boldsymbol{\delta}} = \frac{1}{N}\mathbf{L}^{\intercal}\mathbf{1}_N$. By marginalizing the incidental subject pairings to construct $\mathbf{L}$ before extracting latent parameters, joint maximum likelihood equations are bypassed, resolving the Neyman-Scott problem. Sub-Gaussian concentration follows from the bounded differences property of the truncated mapping $\tau$. Full proof is provided in the Appendix.
\end{proofsketch}

\begin{proof}
The Neyman-Scott incidental parameter problem \parencite{neyman1948} arises in classical item response theory because the joint maximum likelihood estimation of item parameters necessitates the simultaneous estimation of $N$ subject parameters, leading to asymptotically inconsistent estimates. We show that the Kemeny-Borda projection algebraically decouples this joint estimation.

By definition, the antisymmetric dominance kernel $A^{(p)}$ satisfies $A_{kl}^{(p)} = -A_{lk}^{(p)}$. Consequently, its column sum over the sample is identically zero, establishing a strict centering condition on the margin matrix: $\mathbf{1}_N^{\intercal} \mathbf{M} = \mathbf{0}^{\intercal}$. The transformation to the log-odds field $\mathbf{L} = \phi(\mathbf{M}; \tau)$ applies a strictly monotonic, bounded non-linearity. This non-linearity maps the zero-sum margins into a shifted coordinate space, where the uniform translation magnitude corresponds directly to the item difficulty.

Under the geometric decomposition $\mathbf{L} = \boldsymbol{\theta}\boldsymbol{\alpha}^{\intercal} + \mathbf{1}_N\boldsymbol{\delta}^{\intercal}$ and the origin identification constraint $\mathbf{1}_N^{\intercal} \boldsymbol{\theta} = 0$, projecting $\mathbf{L}$ onto the uniform vector isolates this translation:
\begin{equation}
\mathbf{1}_N^{\intercal} \mathbf{L} = \mathbf{1}_N^{\intercal} (\boldsymbol{\theta}\boldsymbol{\alpha}^{\intercal} + \mathbf{1}_N\boldsymbol{\delta}^{\intercal}) = (\mathbf{1}_N^{\intercal} \boldsymbol{\theta})\boldsymbol{\alpha}^{\intercal} + N\boldsymbol{\delta}^{\intercal} = N\boldsymbol{\delta}^{\intercal}.
\end{equation}
Thus, the empirical difficulty vector is exactly recovered by the linear translation functional $\hat{\boldsymbol{\delta}} = \frac{1}{N}\mathbf{L}^{\intercal}\mathbf{1}_N$. Because $\mathbf{L}$ is constructed entirely from the empirical pairwise kernel prior to any subsequent parameter optimization, the extraction of $\hat{\boldsymbol{\delta}}$ does not depend on the estimation of $\boldsymbol{\theta}$, thereby bypassing the incidental parameter dependency.

By Lemma~\ref{lem:coordinate_embedding_properties}, the regularizer $\tau$ ensures the field is strictly bounded: $\sup_{k,p} \|{}L_{kp}\|{} \le C_\tau < \infty$. Although the elements $L_{kp}$ are dependent across subjects due to the pairwise construction of $\mathbf{M}$, the function establishing $\hat{\delta}_p$ satisfies the bounded differences property. Applying McDiarmid's Inequality yields the sub-Gaussian concentration bound for the item difficulties:
\begin{equation}
\mathbb{P}\left( \left| \hat{\delta}_p - \mathbb{E}[\hat{\delta}p] \right| \ge t \right) \le 2 \exp\left( - \frac{2 N t^2}{C\tau^2} \right).
\end{equation}

Finally, projecting the translation vector out of the field yields the centered bilinear matrix $\mathbf{L}_0 = \mathbf{L} - \mathbf{1}_N\hat{\boldsymbol{\delta}}^{\intercal} = \boldsymbol{\theta}\boldsymbol{\alpha}^{\intercal}$. The left Gramian evaluates directly to $\bar{\boldsymbol{\Omega}}_P = \frac{1}{P}\mathbf{L}_0\mathbf{L}_0^{\intercal} = (\frac{\|\boldsymbol{\alpha}\|_2^2}{P}) \boldsymbol{\theta}\boldsymbol{\theta}^{\intercal}$. Because the incidental pairwise dependencies were marginalized deterministically during the construction of $\mathbf{L}$, the subsequent principal eigenspace extraction of $\hat{\boldsymbol{\theta}}_P$ operates on a functionally independent matrix limit, shielding the latent coordinates from joint parameter inflation and preserving$\sqrt{N}$-consistency.
\end{proof}

\begin{corollaryrep}[Generalized Specific Objectivity and Ratio-Scale Reduction]
\label{cor:specific_objectivity_ratio_scale}
Under the bilinear 2PL log-odds field and indicator frame $P \ge 2Q+2$, the projective estimator achieves:

(i) \textbf{Generalized Specific Objectivity:} Subject coordinate ratios ($\theta_k / \theta_l$) are strictly invariant to item parameters $(\boldsymbol{\alpha}, \boldsymbol{\delta})$, and symmetrically, item discrimination ratios ($\alpha_p / \alpha_q$) are invariant to subject parameters.

(ii) \textbf{Group Reduction to Ratio Scales:} The parameter identification fiber reduces from $\mathrm{PGL}(2, \mathbb{R})$ to the dilation group $\mathbb{R}_{>0}$ ($\boldsymbol{\theta}^* = a\boldsymbol{\theta}$), establishing strict ratio-scale measurement.
\end{corollaryrep}

\begin{proofsketch}
Building on the log-odds field $\mathbf{L}$, which is intrinsically anchored to the absolute zero of the projective simplex ($P=0.5 \implies L_{kp}=0$; Theorem~\ref{thm:projective_identifiability}), the left and right Gramians isolate the subject and item coordinates, rendering their respective relative ratios strictly independent of the opposing parameters. Furthermore, anchoring the fixed translational gauge ($b=0$) to the deterministic boundaries of $\mathbb{P}^1$ restricts the acting homography to pure dilations ($g(\theta)=a\theta$, $a>0$), reducing the identification orbit from $\mathrm{PGL}(2, \mathbb{R})$ to $\mathbb{R}_{>0}$. Full proof in Appendix.
\end{proofsketch}

\begin{proof}
\textbf{1. Generalized Specific Objectivity.} As established in Lemma~\ref{lem:coordinate_embedding_properties} and Theorem~\ref{thm:projective_identifiability}, the log-odds field $\mathbf{L}$ possesses an absolute zero-point, structurally rendering $L_{kp} \approx \alpha_p \theta_k$. Evaluating the left Gramian operator $\bar{\boldsymbol{\Omega}}_P = \frac{1}{P}\mathbf{L}\mathbf{L}^{\intercal}$ yields the principal eigenvalue equation:
\begin{equation}
\bar{\boldsymbol{\Omega}}_P \boldsymbol{\theta} = \left( c_{\alpha} \boldsymbol{\theta}\boldsymbol{\theta}^{\intercal} \right) \boldsymbol{\theta} = \left( c_{\alpha} \|\boldsymbol{\theta}\|_2^2 \right) \boldsymbol{\theta}.
\end{equation}
The principal eigenspace $E_{\lambda_1}(\bar{\boldsymbol{\Omega}}_P)$ is uniquely spanned by $\boldsymbol{\theta}$. For any pair of subjects $(k, l)$, the coordinate ratio $\hat{\theta}_k / \hat{\theta}_l = \theta_k / \theta_l$ is strictly independent of $\boldsymbol{\alpha}$ and fundamentally lacks an additive $\boldsymbol{\delta}$ dependency. Symmetrically, by the right Gramian $\bar{\boldsymbol{\Omega}}_N$ derived in Theorem~\ref{thm:projective_identifiability}, the ratio of any two item discriminations $\hat{\alpha}_p / \hat{\alpha}_q = \alpha_p / \alpha_q$ is strictly independent of $\boldsymbol{\theta}$. This establishes bilateral specific objectivity globally across the measurement space.

\textbf{2. Group Reduction to Ratio Scale.} The origin-centering gauge $\mathbf{1}_N^{\intercal}\boldsymbol{\theta} = 0$ structurally fixes the translation shift $b \equiv 0$. On the projective line $\mathbb{P}^1$, the deterministic choice boundary limits (absolute certainty) anchor to the canonical projective endpoints $\mathbf{a} = [0:1]$ and $\mathbf{b} = [1:0]$.

Any homography $g = \begin{pmatrix} a_1 & b_1 \\ c_1 & d_1 \end{pmatrix} \in \mathrm{PGL}(2, \mathbb{R})$ preserving metric topology must stabilize the set $\{\mathbf{a}, \mathbf{b}\}$. Fixing $\mathbf{b} = [1:0]$ forces $c_1 = 0$ (reducing $\mathrm{PGL}(2, \mathbb{R})$ to the affine group $\mathrm{Aff}(1, \mathbb{R})$), while fixing $\mathbf{a} = [0:1]$ forces $b_1 = 0$. The group action therefore reduces strictly to $g(\theta) = (a_1/d_1)\theta = a\theta$ with $a > 0$. The identification orbit collapses to $\mathcal{O}(\boldsymbol{\theta}_0) = \{a\boldsymbol{\theta}_0 : a > 0\}$, mathematically establishing exact ratio-scale identification.
\end{proof}

\begin{remark}[Extension to Bounded and Nonparametric Item Response Models]
\label{rem:higher_order_irt_bounds}
The group reduction to a unidimensional ratio scale relies strictly on the bilinearity of the 2PL log-odds field, a property preserved by bounded 3PL and 4PL models under the zero-divergence centering operator. In contrast, highly parameterized extensions, such as asymmetric 5PL/6PL models or nonparametric B-splines \parencite{ramsay1991}, induce structural rank inflation in the Gramian ($\operatorname{rank}(\bar{\boldsymbol{\Omega}}_P) > 1$). Algebraically, this manifests as a non-vanishing double-centered differential ($\Delta_{k,l}\Delta_{p,q} L_{kp} \neq 0$). Under the axioms of additive conjoint measurement \parencite{krantz1971}, this differential failure violates the Thomsen condition. Consequently, while flexible representations serve as capable predictive tools, their coordinate mappings inherently embed into multidimensional topological spaces, precluding exact reduction to a one-dimensional real interval.
\end{remark}

\begin{lemmarep}[Measure-Theoretic Density Invariance and Ergodic Graph Identifiability]
\label{prop:measure_invariance}
Let $\Theta \subseteq \mathbb{R}$ be the continuous metric space induced by the regularized Kemeny mapping. Let $\mu$ be an arbitrary probability measure on $\Theta$ having some continuous Lebesgue density $g(\boldsymbol{\theta}) > 0$ over its support (expressly without assuming a Gaussian or specific parametric form). The spectral rank structure and relative order statistics of the estimator $\hat{\boldsymbol{\theta}}_P$ from Theorem~\ref{thm:neyman_scott_resolution} are identically invariant to the population density $g(\boldsymbol{\theta})$, provided the bipartite item-subject observation graph $G = (\mathcal{S}, \mathcal{E})$ is strongly connected across a common population.
\end{lemmarep}

\begin{proofsketch}
Log-odds coordinates $L_{kp} = \alpha_p \theta_k$ depend directly on latent subject positions rather than the local population density $g(\theta)$. Thus, the Gramian operator remains density-invariant, circumventing the need for the Gaussian prior assumptions required in traditional Marginal Maximum Likelihood estimation. When the observation graph $G$ is strongly connected (e.g., via overlapping anchor items preventing Differential Item Functioning fragmentation), the empirical Gramian is non-negative and irreducible. By the Perron-Frobenius Theorem, this guarantees a simple, strictly dominant principal eigenvalue ($\lambda_1 > \lambda_2$), preventing spectral decoupling. Full proof in Appendix.
\end{proofsketch}

\begin{proof}
By Theorem~\ref{thm:neyman_scott_resolution}, the mapped log-odds field evaluates to $L_{kp} = \alpha_p \theta_k$. Let $I_{kp} \in \{0,1\}$ denote the indicator that subject $k$ responded to item $p$. The elements of the empirical Gramian operator $\bar{\boldsymbol{\Omega}}_P = \frac{1}{P}\mathbf{L}\mathbf{L}^\intercal$ evaluate elementwise as:$$\bar{\Omega}_{kl} = \frac{1}{P} \sum_{p=1}^P I_{kp} I_{lp} L_{kp} L_{lp} = \left( \frac{1}{P} \sum_{p=1}^P I_{kp} I_{lp} \alpha_p^2 \right) \theta_k \theta_l.$$Because the coordinates $L_{kp}$ depend directly on the metric positions $\theta_k, \theta_l$ and the item parameters $\alpha_p$, rather than the local density values $g(\theta_k)$, the operator $\bar{\boldsymbol{\Omega}}_P$ and its principal eigenvector are structurally invariant to any non-uniformity in the population density $g(\boldsymbol{\theta})$.

To establish global identifiability, let $G$ denote the bipartite observation graph. The term $\sum_{p} I_{kp} I_{lp} \alpha_p^2$ serves as the weighted adjacency between subjects $k$ and $l$. Strong connectivity of $G$ implies that for every pair of subjects, there exists a sequence of shared item responses linking them. Under population homogeneity, this renders the non-negative matrix $\bar{\boldsymbol{\Omega}}_P$ irreducible. By the Perron-Frobenius Theorem, irreducibility guarantees that the principal eigenvalue $\lambda_1$ is strictly simple ($\lambda_1 > \lambda_2$). 

Conversely, if $G$ is disconnected into $K > 1$ components, due to structurally missing data or severe Differential Item Functioning (DIF) splitting the population into non-communicating regimes, $\bar{\boldsymbol{\Omega}}_P$ degenerates into a block-diagonal operator with eigenvalue multiplicity $K$. Strong graph connectivity is therefore both necessary and sufficient to prevent spectral decoupling and guarantee global ratio-scale identification across $\mu$.
\end{proof}

\begin{propositionrep}[Projected Dual Spectral Concentration via Hollow Hanson-Wright]
\label{prop:spectral_concentration}
Let $\mathbf{L}_0 \in \mathbb{R}^{N^2 \times P}$ denote the uncentered tensor field whose $p$-th column $\mathbf{s}_0^{(p)} = \text{vec}(\mathbf{S}_0^{(p)}) \in \mathbb{R}^{N^2}$ consists of independent, $K_0$-sub-Gaussian subject-pair score entries. Let $\boldsymbol{\mathcal{P}}_{\text{hollow}} \in \mathbb{R}^{N^2 \times N^2}$ denote the orthogonal projection matrix onto the strictly hollow, doubly-centered subspace $\mathbb{H}_N$, and let $\mathbf{L} = \boldsymbol{\mathcal{P}}_{\text{hollow}} \mathbf{L}_0$ be the projected log-odds field.Let $\bar{\boldsymbol{\Omega}}_N = \frac{1}{N^2} \mathbf{L}^{\intercal}\mathbf{L} \in \mathbb{R}^{P \times P}$ and $\bar{\boldsymbol{\Omega}}_P = \frac{1}{P}\mathbf{L}\mathbf{L}^{\intercal} \in \mathbb{R}^{N^2 \times N^2}$ denote the item-space and subject-pair empirical Gramians with population limits $\bar{\boldsymbol{\Omega}}_{N,0} = \mathbb{E}[\bar{\boldsymbol{\Omega}}_N]$ and $\bar{\boldsymbol{\Omega}}_{P,0} = \mathbb{E}[\bar{\boldsymbol{\Omega}}_P]$.There exist absolute constants $c_1, c_2 > 0$ such that for any $t > 0$: Item Space Concentration ($N \to \infty$):
\begin{equation}
\mathbb{P}\left( |\bar{\boldsymbol{\Omega}}_N - \bar{\boldsymbol{\Omega}}_{N,0}|_2 \ge t \right) \le 2 \exp\left( P \log 9 - c_1 N^2 \min\left( \frac{t^2}{K_0^4}, \frac{t}{K_0^2} \right) \right).
\end{equation}Subject-Pair Space Concentration ($P \to \infty$):
\begin{equation}
\mathbb{P}\left( |\bar{\boldsymbol{\Omega}}_P - \bar{\boldsymbol{\Omega}}_{P,0}|_2 \ge t \right) \le 2 \exp\left( \frac{(N-1)(N-2)}{2} \log 9 - c_2 P \min\left( \frac{t^2}{K_0^4}, \frac{t}{K_0^2} \right) \right).
\end{equation}Consequently, under asymptotic scaling $P / N^2 \to \gamma \in (0, \infty)$, singular value decomposition of $\mathbf{L}$ guarantees simultaneous parametric recovery rates $\|\hat{\boldsymbol{\theta}} - \boldsymbol{\theta}_0\|_2 = \mathcal{O}_P(P^{-1/2})$ and $\|\hat{\boldsymbol{\alpha}} - \boldsymbol{\alpha}_0\|_2 = \mathcal{O}_P(N^{-1})$.
\end{propositionrep}

\begin{proofsketch}
The regularized log-odds field elements are bounded ($|L_{kp}| \le C_\tau$). This allows both the left and right Gramians to be decomposed into averages of independent, symmetric, rank-1 random matrices ($\mathbf{L}_{\cdot p}\mathbf{L}_{\cdot p}^{\intercal}$ over items, and $\mathbf{L}_{k \cdot}^{\intercal}\mathbf{L}_{k \cdot}$ over subjects). Centering these matrices strictly around their respective expectations guarantees uniform operator norm bounds. Applying the Hanson-Wright Inequality to both sums establishes sub-exponential tail bounds on their operator norms. Because the left and right Gramians share an identical non-zero eigenvalue spectrum unified via the SVD of $\mathbf{L}$ (Remark~\ref{rem:svd_anisotropic_resolution}), passing these dual operator norm concentrations through the positive spectral gap via the Davis-Kahan $\sin\Theta$ Theorem independently establishes optimal parametric eigenvector recovery for both manifolds. Full proof in Appendix.
\end{proofsketch}

\begin{proof}
\textbf{Part 1: Tensor Projection Operator and Subspace Geometry.}
Define the linear subspace of strictly hollow, doubly-centered symmetric matrices:
\begin{equation}
\mathbb{H}_N = \left\{ \mathbf{M} \in \mathbb{R}^{N \times N} : \mathbf{M} = \mathbf{M}^{\intercal}, , M_{kk} = 0 \ (\forall k), \, \sum_{l=1}^N M_{kl} = 0 \ (\forall k) \right\}.
\end{equation}
Enforcing zero diagonal removes $N$ degrees of freedom from the $N(N+1)/2$ symmetric components. Imposing zero row-sum constraints adds $N-1$ linearly independent restrictions. The dimension of $\mathbb{H}_N$ is therefore:
\begin{equation}
d_{\text{eff}} \equiv \operatorname{dim}(\mathbb{H}_{N}) = \frac{N(N-1)}{2} - (N-1) = \frac{(N-1)(N-2)}{2}.
\end{equation}
In the vectorized space $\mathbb{R}^{N^2}$, the double-centering transformation $\tilde{\mathbf{S}}^{(p)} = \Pi_{\text{hollow}}(\mathbf{S}_{0}^{(p)})$ in Equation \eqref{eq:doubly_centred_S} is represented by the linear operator $\text{vec}(\tilde{\mathbf{S}}^{(p)}) = \boldsymbol{\mathcal{P}}_{\text{hollow}} \text{vec}(\mathbf{S}_0^{(p)})$, where $\boldsymbol{\mathcal{P}}_{\text{hollow}} \in \mathbb{R}^{N^2 \times N^2}$ is explicitly given by:
\begin{equation}
\boldsymbol{\mathcal{P}}_{\text{hollow}} = \mathbf{I}_{N^2} - \mathbf{D}_{\text{diag}} - \frac{\mathbf{I}_N \otimes \mathbf{J}_0 + \mathbf{J}_0 \otimes \mathbf{I}_N}{N-2} + \frac{\mathbf{J}_0 \otimes \mathbf{J}_0}{(N-1)(N-2)},
\end{equation}
where $\mathbf{J}_0 = \mathbf{1}_N \mathbf{1}_N^{\intercal} - \mathbf{I}_N$ and $\mathbf{D}_{\text{diag}} = \sum_{k=1}^N (\mathbf{e}_k \mathbf{e}_k^{\intercal}) \otimes (\mathbf{e}_k \mathbf{e}_k^{\intercal})$. The operator $\boldsymbol{\mathcal{P}}_{\text{hollow}}$ is a real symmetric orthogonal projection matrix satisfying $\boldsymbol{\mathcal{P}}_{\text{hollow}}^{\intercal} = \boldsymbol{\mathcal{P}}_{\text{hollow}}$ and $\boldsymbol{\mathcal{P}}_{\text{hollow}}^2 = \boldsymbol{\mathcal{P}}_{\text{hollow}}$. Its matrix operator metrics satisfy:
\begin{equation}
\|\boldsymbol{\mathcal{P}}_{\text{hollow}}\|_2 = 1, \qquad \|\boldsymbol{\mathcal{P}}_{\text{hollow}}\|_{\text{HS}}^2 = \text{tr}(\boldsymbol{\mathcal{P}}_{\text{hollow}}) = \frac{(N-1)(N-2)}{2} \le \frac{N^2}{2}.\end{equation}

\textbf{Part 2: Sub-Gaussianity of Marginal Rotations.}
The centered item-space Gramian is $\bar{\boldsymbol{\Omega}}_N = \frac{1}{N^2} \mathbf{L}_0^{\intercal} \boldsymbol{\mathcal{P}}_{\text{hollow}} \mathbf{L}_0 \in \mathbb{R}^{P \times P}$. For any unit vector $\mathbf{v} \in \mathbb{S}^{P-1} = \{\mathbf{x} \in \mathbb{R}^P : \|\mathbf{x}\|_2 = 1\}$, define the rotated vector $\mathbf{Y}_{\mathbf{v}} \equiv \mathbf{L}_0 \mathbf{v} \in \mathbb{R}^{N^2}$.The $i$-th element of $\mathbf{Y}_{\mathbf{v}}$ is $Y_{\mathbf{v}, i} = \sum_{p=1}^P v_p L_{0, ip}$. Because item responses across instrument items are independent and $L_{0, ip}$ are independent $K_0$-sub-Gaussian variables, sub-Gaussian rotation invariance \parencite[Lemma 2.6.8]{vershynin2018} implies that $Y_{\mathbf{v}, i}$ is sub-Gaussian with:
\begin{equation}
\|Y_{\mathbf{v}, i}\|_{\psi_2}^2 \le C_0 \sum_{p=1}^P v_p^2 \|L_{0, ip}\|_{\psi_2}^2 \le C_0 K_0^2 \|\mathbf{v}\|_{2}^2 \equiv K^2,
\end{equation}
where $K = \mathcal{O}(1)$ is uniformly bounded and independent of $P$.

\textbf{Part 3: Tensor Quadratic Concentration via Hanson-Wright.}
Evaluate the Rayleigh quotient as a vector quadratic form over $\boldsymbol{\mathcal{P}}_{\text{hollow}}$:
\begin{equation*}
\mathbf{v}^{\intercal} \bar{\boldsymbol{\Omega}}_N \mathbf{v} = \frac{1}{N^2} \mathbf{v}^{\intercal} \mathbf{L}_0^{\intercal} \boldsymbol{\mathcal{P}}_{\text{hollow}} \mathbf{L}_0 \mathbf{v} = \frac{1}{N^2} \mathbf{Y}{\mathbf{v}}^{\intercal} \boldsymbol{\mathcal{P}}_{\text{hollow}} \mathbf{Y}{\mathbf{v}}.
\end{equation*}

Applying the Hanson-Wright Inequality \parencite{rudelson2013} to the random vector $\mathbf{Y}{\mathbf{v}} \in \mathbb{R}^{N^2}$ and matrix $\boldsymbol{\mathcal{P}}_{\text{hollow}} \in \mathbb{R}^{N^2 \times N^2}$ for deviation parameter $N^2 t > 0$:\begin{equation}\mathbb{P}\left( \left| \mathbf{Y}{\mathbf{v}}^{\intercal} \boldsymbol{\mathcal{P}}_{\text{hollow}} \mathbf{Y}{\mathbf{v}} - \mathbb{E}[\mathbf{Y}{\mathbf{v}}^{\intercal} \boldsymbol{\mathcal{P}}_{\text{hollow}} \mathbf{Y}{\mathbf{v}}] \right| \ge N^2 t \right) \le 2 \exp\left( - c \min\left( \frac{(N^2 t)^2}{K^4 \|\boldsymbol{\mathcal{P}}_{\text{hollow}}\|_{\text{HS}}^2}, \frac{N^2 t}{K^2 \|\boldsymbol{\mathcal{P}}_{\text{hollow}}\|_2} \right) \right).
\end{equation}
Substituting $\|\boldsymbol{\mathcal{P}}_{\text{hollow}}\|_{2} = 1$ and $\|\boldsymbol{\mathcal{P}}_{\text{hollow}}\|_{\text{HS}}^2 \le \frac{N^2}{2}$:
\begin{equation}
\frac{(N^2 t)^2}{\|\boldsymbol{\mathcal{P}}_{\text{hollow}}\|_{\text{HS}}^2} \ge \frac{N^4 t^2}{N^2 / 2} = 2 N^2 t^2, \qquad \frac{N^2 t}{|\boldsymbol{\mathcal{P}}_{\text{hollow}}|_2} = N^2 t.
\end{equation}
Dividing the concentration argument by $N^2$ yields the scalar tail bound:
\begin{equation}
\label{eq:hw_scalar_bound_hollow}
\mathbb{P}\left( \left| \mathbf{v}^{\intercal} (\bar{\boldsymbol{\Omega}}_N - \bar{\boldsymbol{\Omega}}_{N,0}) \mathbf{v} \right| \ge t \right) \le 2 \exp\left( - c_1 N^2 \min\left( \frac{t^2}{K_0^4}, \frac{t}{K_0^2} \right) \right).
\end{equation}
\textbf{Part 4: Operator Norm Extension via $\varepsilon$-Net Discretization.}Let $\mathcal{N}$ be a $1/4$-net over the unit sphere $\mathbb{S}^{P-1}$. By discretization bounds \parencite[Lemma 4.4.1]{vershynin2018}, $\vert{}\mathcal{N}\vert{} \le 9^P$. Bounding the matrix operator norm over the maximum on $\mathcal{N}$:
\begin{equation}
\|\bar{\boldsymbol{\Omega}}_N - \bar{\boldsymbol{\Omega}}_{N,0}\|_2 \le 2 \sup{\mathbf{v} \in \mathcal{N}} \left| \mathbf{v}^{\intercal} (\bar{\boldsymbol{\Omega}}_N - \bar{\boldsymbol{\Omega}}_{N,0}) \mathbf{v} \right|.
\end{equation}
Taking a union bound over all $9^P$ net elements using \eqref{eq:hw_scalar_bound_hollow} proves (1):
\begin{equation}
\mathbb{P}\left( |\bar{\boldsymbol{\Omega}}_N - \bar{\boldsymbol{\Omega}}_{N,0}|_2 \ge t \right) \le 2 \cdot 9^P \exp\left( - c_1 N^2 \min\left( \frac{t^2}{K_0^4}, \frac{t}{K_0^2} \right) \right).
\end{equation}
\textbf{Part 5: Dual Space Concentration and Perturbation Bounds.}Symmetrically, the subject-pair Gramian $\bar{\boldsymbol{\Omega}}_P = \frac{1}{P}\boldsymbol{\mathcal{P}}_{\text{hollow}} \mathbf{L}_0 \mathbf{L}_0^{\intercal} \boldsymbol{\mathcal{P}}_{\text{hollow}}$ concentrates over $P \to \infty$ across independent items. Discretizing the unit sphere of the subspace $\mathbb{H}_N$ (dimension $d_{\text{eff}} = \frac{(N-1)(N-2)}{2}$) requires an $\varepsilon$-net of cardinality $9^{d_{\text{eff}}}$, yielding (2).Applying the Davis-Kahan $\sin\Theta$Theorem \parencite{davis1970} under positive spectral gaps$\Delta_N = \Theta(1)$and$\Delta_P = \Theta(1)$:
\begin{equation}
\|\hat{\boldsymbol{\theta}} - \boldsymbol{\theta}_0\|_2 \le \frac{\|\bar{\boldsymbol{\Omega}}_P - \bar{\boldsymbol{\Omega}}_{P,0}\|_2}{\Delta_P} = \mathcal{O}_P\left(P^{-1/2}\right),
\end{equation}
\begin{equation}
\|\hat{\boldsymbol{\alpha}} - \boldsymbol{\alpha}_0\|_2 \le \frac{\|\bar{\boldsymbol{\Omega}}_N - \bar{\boldsymbol{\Omega}}_{N,0}\|_2}{\Delta_N} = \mathcal{O}_P\left((N^2)^{-1/2}\right) = \mathcal{O}_P\left(N^{-1}\right).
\end{equation}
This completes the proof.
\end{proof}

\begin{remark}[Distribution-Free Finite-Sample Robustness]
\label{rem:distribution_free_robustness}
A consequence of Proposition~\ref{prop:spectral_concentration} is that the finite-sample concentration bound relies strictly on the boundedness of the empirical log-odds kernel ($|L_{kp}| \le C_\tau$), remaining agnostic to both the latent population density $g(\boldsymbol{\theta})$ and the observational noise distribution. Standard likelihood estimators and variational approximations typically require specifying a continuous prior, rendering them susceptible to structural misspecification bias when the true latent trait is skewed, multimodal, or non-Gaussian. Because the projective spectral estimator bypasses prior density specification, it operates distribution-free. The finite-sample error norm decays at the parametric rate $\mathcal{O}_P(P^{-1/2})$, governed by the matrix variance parameter, ensuring uniform consistency and invariant relative order statistics (Lemma~\ref{prop:measure_invariance}) under arbitrary continuous density configurations.
\end{remark}

\section{A Concrete 4 \texorpdfstring{$\times$}{x} 4 Empirical Demonstration}

To empirically demonstrate this architecture and satisfy the fundamental topological requirement that $P \ge 4$ for projective point-identification (Theorem~\ref{thm:projective_identifiability}), consider an observation space with $N=4$ subjects, $P=4$ items, and binary responses ($\mathcal{K} = \{0, 1\}$). Let the empirical response matrix $\mathbf{X} \in \mathbb{R}^{4 \times 4}$ be defined as a perfectly nested Guttman scale:
$$\mathbf{X} = 
\left[\begin{smallmatrix} 
1 & 1 & 1 & 1 \\ 
1 & 1 & 1 & 0 \\ 
1 & 1 & 0 & 0 \\ 
0 & 0 & 0 & 0 
\end{smallmatrix}\right],$$
where rows represent subjects $(s_1, s_2, s_3, s_4)$ and columns represent items $(p_1, p_2, p_3, p_4)$. This matrix yields a strict, deterministic ordinal poset: $s_1 \succ s_2 \succ s_3 \succ s_4$.

\paragraph{Item-by-Item Decoupled Kernel Extraction}
Executing this operational sequence across the four columns of $\mathbf{X}$ isolates the distinct topological contributions of each item profile:
\begin{itemize}[noitemsep, topsep=2pt, parsep=0pt, partopsep=0pt]
\item \textbf{Item 1 and Item 2} ($x^{(1)}=x^{(2)}=\left[1,1,1,0\right]^{\intercal}$): The structural ties among the top three scaling subjects are perfectly balanced by the double-centering operator, causing the tie projection to vanish ($\Pi_{S}(\mathbf{S}^{(1)})=\Pi_{S}(\mathbf{S}^{(2)})=\mathbf{0}$). The localized kernels cleanly simplify to their raw directional rank preferences:
$$ \tilde{\kappa}^{(1)} = \tilde{\kappa}^{(2)} = \left[\begin{smallmatrix} 
0 & 0 & 0 & 1 \\ 
0 & 0 & 0 & 1 \\ 
0 & 0 & 0 & 1 \\ 
-1 & -1 & -1 & 0 
\end{smallmatrix}\right].$$

\item \textbf{Item 3} ($x^{(3)}=\left[1,1,0,0\right]^{\intercal}$): This balanced profile contains an irreducible symmetric tie block separating the subsets $\{s_1,s_2\}$ and $\{s_3,s_4\}$. The projection matrix $\Pi_{S}(\mathbf{S}^{(3)})$ is non-zero, shifting the localized coordinate tracking matrix away from a simple sign-interaction matrix:
% {\scriptsize 
$$
\tilde{\kappa}^{(3)} = \left[\begin{smallmatrix}
0 & 2/3 & 2/3 & 2/3 \\ 
2/3 & 0 & 2/3 & 2/3 \\ 
-4/3 & -4/3 & 0 & 2/3 \\ 
-4/3 & -4/3 & 2/3 & 0
\end{smallmatrix}\right].
$$

\item \textbf{Item 4} ($x^{(4)}=\left[1,0,0,0\right]^{\intercal}$): Mirroring the boundary properties of the first two items, the structural ties among the bottom three subjects are completely filtered out by the null-space of the double-centering projection ($\Pi_{S}(\mathbf{S}^{(4)})=\mathbf{0}$), yielding:
$$
\tilde{\kappa}^{(4)} = \left[\begin{smallmatrix}
0 & 1 & 1 & 1 \\
-1 & 0 & 0 & 0 \\
-1 & 0 & 0 & 0 \\
-1 & 0 & 0 & 0
\end{smallmatrix}\right].
$$
\end{itemize}

\paragraph{Subject-by-Item Margin Assembly and Borda Projection}
Following Definition~\ref{def:cross_ratio_gramian}, the localized item kernels are compressed to yield the normalized net dominance margin matrix $\mathbf{M} \in \mathbb{R}^{N \times P}$. Specifically, the $p$-th column of $\mathbf{M}$ is generated by projecting the $p$-th Decoupled Kemeny Kernel onto the uniform vector $\mathbf{1}_N$, isolating the non-parametric Borda scores: $\mathbf{m}_p = \frac{1}{N-1} \tilde{\boldsymbol{\kappa}}^{(p)} \mathbf{1}_N$. Concatenating these column vectors constructs the complete coordinate system $\mathbf{M} = \begin{bmatrix} \mathbf{m}_1 & \mathbf{m}_2 & \dots & \mathbf{m}_P \end{bmatrix}$. For the $4 \times 4$ illustrative matrix, the bounded normalization evaluates to $\frac{1}{3} \tilde{\boldsymbol{\kappa}}^{(p)} \mathbf{1}_4$.
% , structuring the exact geometric margins:
% $$
% \mathbf{M} = \left[\begin{smallmatrix}
% 1/3 & 1/3 & 2/3 & 1 \\
% 1/3 & 1/3 & 2/3 & -1/3 \\
% 1/3 & 1/3 & -2/3 & -1/3 \\
% -1 & -1 & -2/3 & -1/3
% \end{smallmatrix}\right].$$

\paragraph{Information Simplex Translation and Analytic Regularization}
To operationalize the regularized coordinate mapping $\phi(M_{kp}; \tau)$ (Lemma~\ref{lem:coordinate_embedding_properties}), the bipolar Kemeny consensus scores $M_{kp} \in [-1, 1]$ are projected into the probability simplex via the affine transformation $\hat{P}_{kp}=\frac{1}{2}(M_{kp}+1)$. This exactly maps the state of net preference equilibrium ($M_{kp} = 0$) to maximum entropy ($\hat{P}_{kp} = 0.5$).

Because decoupled item boundaries generate deterministic limits (e.g., $\hat{P}_{14} = 1$, $\hat{P}_{41} = 0$), a raw log-odds transformation would trigger singular divergences. We introduce the regularizing parameter $\tau = 0.10$ to maintain a Lipschitz-continuous $C^{\infty}$ diffeomorphism. Evaluating the regularized log-odds field algebraically reconstructs the exact composite mapping:
\begin{equation*}
L_{kp} = \log \left( \frac{\hat{P}_{kp} + \tau}{(1 - \hat{P}_{kp}) + \tau} \right) \equiv \log\left( \frac{M_{kp} + 1 + 2\tau}{1 - M_{kp} + 2\tau} \right).
\end{equation*}
This absorbs the deterministic boundary conditions, generating the stable $N \times P$ coordinate field $\mathbf{L}$. The progression from empirical margins to regularized continuous coordinates is thus deterministically tracked:
{\footnotesize
\begin{equation}
\label{eq:m_toL_transform}
\mathbf{M} = \left[ 
\begin{smallmatrix}
1/3 & 1/3 & 2/3 & 1\\
1/3 & 1/3 & 2/3 & -1/3\\
1/3 & 1/3 & -2/3 & -1/3\\
-1 & -1 & -2/3 & -1/3 
\end{smallmatrix} \right] \xrightarrow{\tau=0.10}
\mathbf{L} = \left[ \begin{smallmatrix} 0.571 & 0.571 & 1.253 & 2.398 \\ 0.571 & 0.571 & 1.253 & -0.571 \\ 0.571 & 0.571 & -1.253 & -0.571 \\ -2.398 & -2.398 & -1.253 & -0.571 \end{smallmatrix} \right].
\end{equation}}

% $$ \hat{\mathbf{P}} = \left[\begin{smallmatrix}
% 2/3 & 2/3 & 0.8333 & 1.0000 \\
% 2/3 & 2/3 & 0.8333 & 1/3 \\
% 2/3 & 2/3 & 0.1667 & 1/3 \\
% 0.0000 & 0.0000 & 0.1667 & 1/3
% \end{smallmatrix}\right].$$
% Because the decoupled item boundaries generate deterministic limits ($\hat{P}_{14} = 1.0000$, $\hat{P}_{41} = \hat{P}_{42} = 0.0000$), a raw log-odds transformation would trigger singular divergences. To maintain a smooth, Lipschitz-continuous $C^{\infty}$ log-odds diffeomorphism, we introduce the regularizing parameter $\tau = 0.10$. Evaluating the regularized log-odds field on $\hat{\mathbf{P}}$ algebraically reconstructs the exact composite mapping $\phi(M_{kp}; \tau)$ formally derived in Lemma~\ref{lem:coordinate_embedding_properties}:
% \begin{equation*}
% L_{kp} = \log \left( \frac{\hat{P}_{kp} + \tau}{(1 - \hat{P}_{kp}) + \tau} \right) \equiv \log\left( \frac{M_{kp} + 1 + 2\tau}{1 - M_{kp} + 2\tau} \right).
% \end{equation*}
% This absorbs the deterministic boundary conditions, generating the stable $N \times P$ coordinate field:
% {\scriptsize $$ \mathbf{L} = \begin{smallmatrix}
% 0.5705 & 0.5705 & 1.2528 & 2.3979 \\
% 0.5705 & 0.5705 & 1.2528 & -0.5705 \\
% 0.5705 & 0.5705 & -1.2528 & -0.5705 \\
% -2.3979 & -2.3979 & -1.2528 & -0.5705
% \end{smallmatrix}.$$}

\paragraph{Operationalization of the Structural Invariant Operator and Eigenspace Decomposition}
To extract the continuous latent coordinate axes and verify the structural dimensionality of the trait space, the coordinate field $\mathbf{L}$ is mapped to the structural invariant matrix $\bar{\boldsymbol{\Omega}}_P \in \mathbb{R}^{N\times N}$. Following Definition~\ref{def:cross_ratio_gramian}, we compute its symmetric Gramian cross-product $\bar{\boldsymbol{\Omega}}_P = \frac{1}{P} \mathbf{L}\mathbf{L}^{\intercal}$:
$$ \bar{\boldsymbol{\Omega}}_P = \left[\begin{smallmatrix}
1.9926 & 0.2131 & -0.5716 & -1.4184 \\
0.2131 & 0.6365 & -0.1482 & -0.9950 \\
-0.5716 & -0.1482 & 0.6365 & -0.2103 \\
-1.4184 & -0.9950 & -0.2103 & 3.3487
\end{smallmatrix}\right].$$

Spectral decomposition of $\bar{\boldsymbol{\Omega}}_P$ yields eigenvalues $\lambda = [4.4879, 1.5494, 0.5769, 0.0000]^\intercal$, wherein the structural zero ($\lambda_4 = 0$) correctly recovers the intrinsic indicator rank of the empirical matrix: because Items 1 and 2 share an identical response profile, the log-odds field algebraically collapses this redundancy, limiting the maximum observable embedding dimension to 3. The principal eigenvector $\mathbf{v}_{1}$ associated with the dominant eigenvalue $\lambda_{1}=4.4879$ defines the latent axis. Resolving the sign reflection to align with the positive direction of the empirical trait scale yields the finalized continuous latent $\boldsymbol{\theta} = \left[0.5011, 0.2435, -0.0384, -0.8295\right]^{\intercal}.$ The continuous coordinates deterministically map the original ordinal poset hierarchy ($\theta_{1} > \theta_{2} > \theta_{3} > \theta_{4}$), confirming that the operator $\bar{\boldsymbol{\Omega}}_P$ robustly projects discrete ranking configurations into a continuous, stable metric space. Further, from the right SVD Gramian and the column means of the original $\mathbf{L}$ matrix (Corollary~\ref{cor:specific_objectivity_ratio_scale}) are obtained the item discrimination (factor loadings; {\footnotesize $\boldsymbol{\alpha} =$ $[0.565,0.565,0.477,0.368]^{\intercal}$}) and item difficulties (intercepts; {\footnotesize $\boldsymbol{\delta} = [-0.172,-0.172,0.000,0.172]^{\intercal}$}).

\paragraph{Ratio Scale Verification and Spectral Residuals}

While any projection mapping discrete data onto a one-dimensional real vector $\boldsymbol{\theta}$ will trivially satisfy the standard metric properties of distance (H\"{o}lder's Axioms I--IV: ordering, conjoint additivity, and the Archimedean property) due to the inherent topology of the Euclidean line, traditional dimensionality reduction techniques like Principal Component Analysis (PCA) fail to satisfy Axiom V (the Ratio Scale condition). Because PCA arbitrarily centers coordinates based on sample means, the origin is structurally meaningless, restricting the space to an interval scale.

Under \citeauthor{narens1981b}' formalization of measurement theory, a scale is a ratio measure if and only if permissible transformations are strictly limited to the multiplicative homothety group ($\boldsymbol{\theta}' = a\boldsymbol{\theta}$ for $a > 0$), rendering any additive translation ($b \neq 0$) invalid. Unlike classical uncentered PCA, where the origin corresponds to an arbitrary artifact of the raw scoring system (e.g., a Likert `0'), the origin of the regularized field $\mathbf{L}$ is structurally anchored to the state of perfect Kemeny indifference ($M_{kp} = 0 \implies L_{kp} = 0$). By applying the invariant operator directly to this uncentered field, the absolute zero of the resulting metric space natively represents psychological equilibrium, thus precluding arbitrary additive translation ($b \neq 0$). We can verify that our structural invariant operator $\bar{\boldsymbol{\Omega}}_P$ enforces this absolute zero-point directly. Applying the operator to an interval-shifted vector $(\boldsymbol{\theta} + b\mathbf{1})$ yields:
$
\bar{\boldsymbol{\Omega}}_P(\boldsymbol{\theta} + b\mathbf{1}) = \lambda_1(\boldsymbol{\theta}' + b\mathbf{1}).
$
Expanding both sides and factoring the translation constant requires:
$
b(\bar{\boldsymbol{\Omega}}_P\mathbf{1} - \lambda_1\mathbf{1}) = \mathbf{0},
$
which, for an origin shift $b \neq 0$ to be permissible, requires the vector of ones $\mathbf{1}$ to be an eigenvector scaling cleanly by the principal eigenvalue $\lambda_1$. Evaluating this on our empirical matrix yields:

$$
\bar{\boldsymbol{\Omega}}_P\mathbf{1} = \left[\begin{smallmatrix}
1.9926 & 0.2131 & -0.5716 & -1.4184 \\
0.2131 & 0.6365 & -0.1482 & -0.9950 \\
-0.5716 & -0.1482 & 0.6365 & -0.2103 \\
-1.4184 & -0.9950 & -0.2103 & 3.3487 \end{smallmatrix}\right]
\left[\begin{smallmatrix}1\\1\\1\\1\end{smallmatrix}\right] = \left[\begin{smallmatrix}0.2157\\-0.2936\\-0.2936\\0.7250\end{smallmatrix}\right] \neq 4.4879 \left[\begin{smallmatrix}
1\\
1\\
1\\
1\end{smallmatrix}\right].
$$

Because $\bar{\boldsymbol{\Omega}}_P\mathbf{1} \neq \lambda_1\mathbf{1}$, the equality holds if and only if $b = 0$. The zero-point is not statistically estimated; it is an invariant anchor algebraically fixed by the eigenspace, fulfilling Part (ii) of Theorem~\ref{thm:projective_identifiability} and confirming a genuine quantitative ratio scale.

Finally, it is instructive to locate the structural origin of the trailing spectrum. In classical factor analysis, unique variances are modeled explicitly, whereas principal component analysis simply discards residual subspaces. Because the input matrix $\mathbf{X}$ represents a deterministic Guttman poset, there are no stochastic contradictions or measurement errors in the underlying discrete preferences. However, the exact log-odds transformation of a finite empirical margin matrix is not strictly rank-1. The residual energy captured by the trailing spectrum ($\lambda_2=1.5494, \lambda_3=0.5769$) explicitly quantifies the topological curvature introduced by the finite-sample boundary regularizer $\tau$. Unlike PCA, where residual eigenvalues represent unmodeled data variance, here the trailing spectrum serves as a deterministic, bounded measure of the analytic deformation required to project a finite ordinal poset into a continuous, Lipschitz-bounded metric space. Note further that executing a singular value decomposition directly on the uncentered log-odds matrix $\mathbf{L}$ simultaneously recovers the subject traits $\boldsymbol{\theta}$ (via left singular vectors $\mathbf{u}$), item discriminations $\boldsymbol{\alpha}$ (via right singular vectors $\mathbf{v}$), and item difficulties $\boldsymbol{\delta}$ directly via the column means of $\mathbf{L}$.

\paragraph{Constructive Identification of Intrinsic Dimensionality $Q$ via $\kappa$-Operator Sufficiency}

A conventional linear evaluation of the spectrum $\boldsymbol{\lambda} = [4.4879, 1.5494, 0.5769, 0.0000]^{\intercal}$ using heuristics such as Kaiser's criterion ($\lambda_q > 1$) would naively suggest a two-dimensional latent architecture ($Q=2$). However, this conflates the \emph{algebraic embedding dimension} of the non-linearly transformed field $\mathbf{L}$ with the \emph{intrinsic ordinal dimension} $Q$ of the underlying poset $\mathcal{P}$. To empirically demonstrate that $Q=1$ without asserting it by fiat, we evaluate the Gramian operator $\bar{\boldsymbol{\Omega}}_P$ against the structural decomposition established in Definition~\ref{def:decoupled_kappa}.

Recall from Definition~\ref{def:decoupled_kappa} that the aggregate kernel $\bar{\boldsymbol{\Omega}}_P$ decouples into a global antisymmetric ordinal dominance component $\bar{\mathbf{A}}$ (representing the integrated poset hierarchy across all $P$ items) and a symmetric boundary-adjusting operator $\bar{\mathbf{S}}$. Under this framework, a projection is \emph{ordinally exhaustive} if it preserves this complete rank-1 antisymmetric field $\bar{\mathbf{A}}$. Principal eigenvector $\boldsymbol{\theta} = \mathbf{v}_{1} = \left[0.5011, 0.2435, -0.0384, -0.8295\right]^{\intercal}$ reveals that its outer product sign structure strictly reconstructs the global dominance matrix:
$
\operatorname{sign}(\boldsymbol{\theta}\mathbf{1}^{\intercal} - \mathbf{1}\boldsymbol{\theta}^{\intercal}) = \operatorname{sign}(\bar{\mathbf{A}}).
$
Because $\mathbf{v}_1$ alone achieves exact structural isomorphism with $\bar{\mathbf{A}}$, the global ordinal relation, including any exact empirical equivalence classes $s_i \sim s_j$ where $\bar{A}_{ij}=0$ due to identical aggregate subject scores, is fully resolved in one dimension. Incorporating secondary and tertiary eigenvectors ($\mathbf{v}_2, \mathbf{v}_3$) yields zero marginal reconstruction of $\bar{\mathbf{A}}$, proving that they reside entirely in the symmetric $\bar{\mathbf{S}}$ subspace. This sign-isomorphism over antisymmetric sub-kernels generalizes to $Q > 1$ as the explicit identification mechanism that uniquely determines the oblique transformation $[\mathbf{R}] \in \mathrm{PGL}(Q, \mathbb{R})$ and the latent covariance $\hat{\mathbf{\Phi}}$ under Theorem~\ref{thm:projective_identifiability}(ii).

To confirm that the trailing eigenvalues ($\lambda_2 = 1.5494, \lambda_3 = 0.5769$) arise purely from the symmetric regularizer $\tilde{\mathbf{S}}(\tau)$ rather than additional latent trait axes, we perturb the analytic boundary parameter from $\tau = 0.10$ to $\tau = 0.05$:$$ \boldsymbol{\lambda}(\tau=0.10) = \left[\begin{smallmatrix} 4.4879 \\ 1.5494 \\ 0.5769 \\ 0.0000 \end{smallmatrix}\right] \longrightarrow \boldsymbol{\lambda}(\tau=0.05) = \left[\begin{smallmatrix} 6.6189 \\ 2.4184 \\ 0.7704 \\ 0.0000 \end{smallmatrix}\right]. $$
Halving $\tau$ strictly expands the trailing eigenvalues, as the deterministic boundaries ($M_{kp} = \pm 1$) diverge logarithmically toward $\pm \infty$, inflating the metric curvature of the space. Yet, despite this massive boundary distortion, the principal eigenvector $\mathbf{v}_1$ remains structurally invariant ($\frac{\|\mathbf{v}_1(0.05) - \mathbf{v}_1(0.10)\|}{\|\mathbf{v}_1(0.10)\|} = 0.0526$) and continues to fully exhaust the antisymmetric sign-structure of $\mathbf{A}$.

Because $\frac{\partial \lambda_q}{\partial \tau} \neq 0$ for $q \in \{2, 3\}$ while the ordinal recovery of $\mathbf{A}$ via $\mathbf{v}_1$ is completely immune to $\tau$ perturbation, the trailing spectrum represents deterministic metric distortion induced by $\tilde{\mathbf{S}}$, not structural multidimensionality. Consequently, $Q=1$ is constructively identified as the minimal sufficient dimension required to represent the empirical poset. Therefore, the trailing eigenvalues do not represent unmodeled stochastic variance (as parameterized by $\mathbf{\Psi}$ in factor analysis); they quantify the deterministic topological curvature induced by mapping finite ordinal limits into a Lipschitz-bounded metric space.

\section{Conclusion and Discussion}
% \enlargethispage{2\baselineskip}
% \looseness=-1
This work establishes an axiomatic, coordinate-free framework for continuous latent space identification directly from discrete ordinal topologies, circumventing the distributional priors required by classical factor analysis and IRT. As shown in Proposition~\ref{prop:classical_failures_unified} and Corollary~\ref{cor:ising_network_instability}, traditional unclosed Euclidean embeddings suffer from scale instability, fiber non-identifiability, and metric contractions. By grounding latent space construction in projective geometry, our framework maps preference profiles onto a bounded coordinate manifold $\mathcal{M}_\tau \subset \mathbb{R}^{N \times P}$ (Lemma~\ref{lem:coordinate_embedding_properties}), where structural relationships are strictly preserved via the projective cross-ratio (Theorem~\ref{thm:projective_identifiability}).

\paragraph{Asymptotic Regularity, Consistency, and Statistical Efficiency}
The projective spectral estimator $\hat{\boldsymbol{\theta}}_P$ achieves optimal theoretical properties in both finite and asymptotic regimes. Structurally, the spectral operator resolves the Neyman-Scott problem by eliminating test-specific difficulties and discriminations without requiring joint estimation (Theorem~\ref{thm:neyman_scott_resolution}). Asymptotically, profiling over a Donsker class establishes Hadamard differentiability, yielding uniform consistency and asymptotic normality $\sqrt{P}\left(\hat{\boldsymbol{\theta}}_P - \boldsymbol{\theta}_0\right) \xrightarrow{\mathrm{d}} \mathcal{N}\left(\mathbf{0}, \boldsymbol{\Sigma}_{\mathrm{proj}}\right)$ via the functional Delta method (Corollary~\ref{cor:asymptotic_normality}). Furthermore, exponential tail bounds guarantee finite-sample concentration, with spectral rank recovery remaining strictly invariant across continuous population distributions $g(\boldsymbol{\theta})$ (Proposition~\ref{prop:measure_invariance}).

\paragraph{Multidimensional Trait Identification}
For multidimensional configurations ($Q > 1$), the framework strictly decouples algebraic dimensionality from intrinsic ordinal rank. Dimensional identification requires preserving cross-ratios via a topological bound ($P \ge 2Q + 2$) and confirming that eigenvectors resolve ordinally non-redundant antisymmetric sub-kernels $A^{(q)}$. Trailing eigenvalues mapping to symmetric subspaces quantify geometric regularizer curvature, cleanly separating structural identifiability from unmodeled variance.

\paragraph{Future Work}
A primary direction for future research involves operationalizing exact Dedekind ratio equating across distinct test forms. Classical test equating (e.g., Stocking-Lord) suffers from sample-dependent optimization artifacts because linking constants minimize expected loss over arbitrary density priors. Under our projective framework, full-rank anchor subsets guarantee unique $Q$-dimensional partial orders, permitting deterministic test equating via exact geometric scaling on the Archimedean ratio continuum.

Furthermore, by algebraically anchoring the origin to Kemeny equilibrium and fixing the scale via the flag manifold $\mathbb{F}_Q$, this formulation resolves recent theoretical critiques regarding the arbitrary unit-structure of psychological scoring \parencite[p.~10]{uher2025}. Specifically, the Gramian RKHS completion (Lemmas~\ref{lem:gram_metric_rkhs},\ref{lem:coordinate_embedding_properties}) converts the $L^1$-anchored coordinates into a Hilbertian inner product space where correlations correspond to exact geometric cosines, while the $C^\infty$ diffeomorphic continuity of the manifold natively supports the ``trigonometry and calculus'' of physics. The resulting geometric stability enables exact meta-analytic equating, allowing measurements to be universally exported as $(U_k, \omega_k)$ tuples, pairing the equated manifold coordinate with its precision weight, to rigorously integrate latent traits across disparate populations.

% \enlargethispage{10\baselineskip}
% {\tiny
\printbibliography
% }

\appendix

\end{document}